%% file: main.tex
\documentclass[5p,times]{elsarticle}

\usepackage{ecrc_RIAI}

\usepackage[english]{babel}     
\addto\captionsenglish{%
}
\UseRawInputEncoding
\usepackage{amsmath}            
\usepackage{epstopdf}           
\usepackage{flushend}           

\volume{00}

\firstpage{1}

\journalname{Biomedical Signal Processing and Control}

\jid{SAP}

\jnltitlelogo{}

\usepackage[figuresright]{rotating}

\usepackage{amssymb}

\runauth{X. Li et al.}

\usepackage{adjustbox}
\usepackage{seqsplit}
\usepackage{booktabs}
\usepackage{algorithm,algorithmic}
\usepackage{xcolor}
\usepackage{multirow}
\usepackage{graphicx}
\usepackage{svg}
\usepackage{url} 
\usepackage{float}
\usepackage{subcaption}

\begin{document}

\begin{frontmatter}

\title{DCentNet: Decentralized Multistage Biomedical \\
Signal Classification using Early Exits}

\author[First]{Xiaolin Li\corref{cor1}}
\ead{xiaolin.li@ucdconnect.ie}

\author[First]{Binhua Huang}
\ead{binhua.huang@ucdconnect.ie}

\author[First]{Barry Cardiff}
\ead{barry.cardiff@ucd.ie}

\author[First]{Deepu John}
\ead{deepu.john@ucd.ie}

\cortext[cor1]{Corresponding author.}

\address[First]{School of Electrical and Electronics Engineering, University College Dublin, Ireland.}

\input{Sections/abstract}

\end{frontmatter}

\input{Sections/1_intro}
\input{Sections/2_relatedwork}
\input{Sections/3_model}
\input{Sections/3.3_EmbedSys}
\input{Sections/4_dataset}
\input{Sections/5_results}

\input{Sections/6_conclusion}

\section*{Acknowledgment}
This work was supported in part by 1) the China Scholarship Council, 2) the Microelectronic Circuits Centre Ireland, and 3) the Irish Research Council.

\bibliographystyle{elsarticle-num}
\bibliography{references}

\end{document}

%% file: Sections/abstract.tex
\begin{abstract}
This paper presents DCentNet, a novel decentralized multistage signal classification approach for biomedical data obtained from Internet of Things (IoT) wearable sensors, utilizing early exit point (EEP) to improve both energy efficiency and processing speed. Traditionally, IoT sensor data is processed in a centralized manner on a single node, Cloud-native or Edge-native, which comes with several restrictions, such as significant energy consumption on the edge sensor and greater latency. To address these limitations, we propose DCentNet, a decentralized method based on Convolutional Neural Network (CNN) classifiers, where a single CNN model is partitioned into several sub-networks using one or more EEPs. Our method introduces encoder-decoder pairs at EEPs, which serve to compress large feature maps before transferring them to the next sub-network, drastically reducing wireless data transmission and power consumption. When the input can be confidently classified at an EEP, the processing can terminate early without traversing the entire network. To minimize sensor energy consumption and overall complexity, the initial sub-networks can be set up in the fog or on the edge. We also explore different EEP locations and demonstrate that the choice of EEP can be altered to achieve a trade-off between performance and complexity by employing a genetic algorithm approach. DCentNet addresses the limitations of centralized processing in IoT wearable sensor data analysis, offering improved efficiency and performance. The experimental results of electrocardiogram (ECG) classification validate the success of our proposed method. With one EEP, the system saves 94.54\% of wireless data transmission and a corresponding 21\% decrease in complexity, while the classification accuracy and sensitivity remain almost unaffected and stay at their original levels. When employing two EEPs, the system demonstrates a sensitivity of 98.36\% and an accuracy of 97.74\%, concurrently leading to a 91.86\% reduction in wireless data transmission and a reduction in complexity by 22\%. DCentNet is implemented on an ARM Cortex-M4 based microcontroller unit (MCU). In laboratory testing, our approach achieves an average power saving of 73.6\% compared to continuous wireless transmission of ECG signals. 
\end{abstract}

\begin{keyword}
Decentralized Inferencing, DNN partitioning, Early Exits, Biomedical Signal Classification, Internet of Things (IoT), Internet of Medical Things (IoMT), Arrhythmia, ECG Classification
\end{keyword}

%% file: Sections/1_intro.tex
\section{Introduction}
\label{intro}
Arrhythmia is characterized by an abnormality in the rhythm of the heart, leading to irregular contractions or a rhythm that is either too slow or too fast, consequently impeding the heart's ability to pump blood and diminishing the efficiency of its blood supply. Inadequate blood supply to vital organs and tissues of the body from arrhythmias can culminate in a myriad of grave consequences, such as heart failure, stroke, and myocardial infarction. ECG investigations are extremely important in the diagnosis of heart arrhythmias. ECG signal analysis, being a prevalent method for detecting arrhythmias, allows for the assessment of individual waveforms and characteristics of ECG signals, thereby enabling the determination of heart health status and identification of potential arrhythmic types. Given the increasing prevalence of wearable health-monitoring devices, the ability to continuously and efficiently monitor ECG signals is crucial for early detection and prevention of cardiovascular events, which are leading causes of mortality worldwide. However, existing systems, which typically rely on centralized model deployment (either entirely on the cloud or on the edge) face challenges in terms of energy efficiency, latency, and computational resources, which this research addresses. Consequently, proper classification and identification of ECG signals can facilitate the early detection of arrhythmias and facilitate the implementation of appropriate therapeutic measures, ultimately enhancing treatment efficacy and reducing patient risk.

\begin{figure*}[ht]
    \centering
    \includegraphics[width=\linewidth]{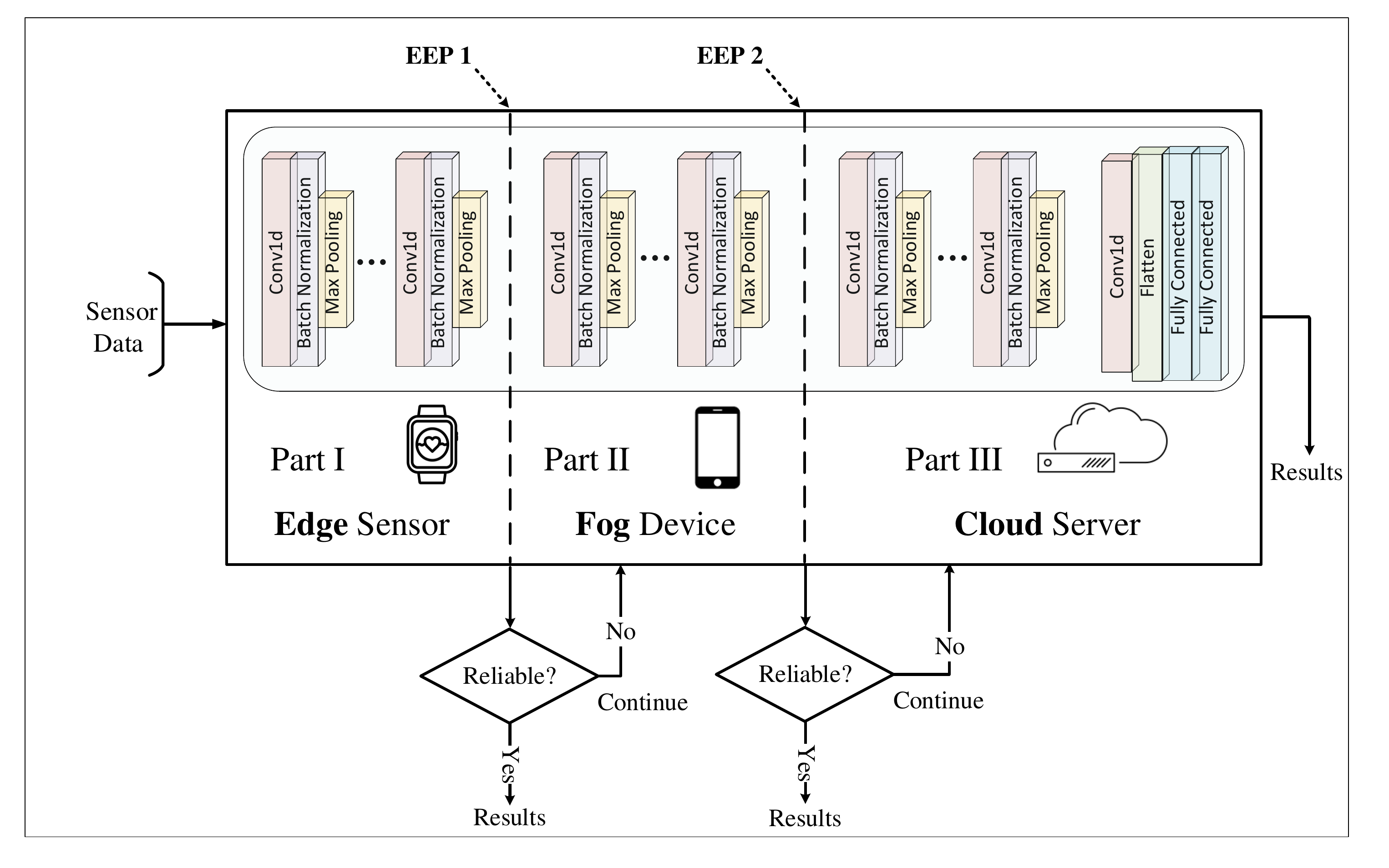}
    \caption{Illustration of the decentralized multistage biomedical inferencing using Early Exits.}
    \label{fig:diagram}
\end{figure*}

Because of the rapid progress of wearable biomedical sensor technologies, it is now possible to capture extended hours of continuous biosignal data, such as ECG signals. Traditionally, data collected by wearable devices is sent to an intermediate gateway before being sent to a cloud server for analysis. For continuous and long-term ECG data analysis, cloud-native applications with automated data processing algorithms are used. CNNs have shown encouraging results in the processing of ECG signals~\cite{xiaolin20201d, SIVA_TBCas}. However, because of the continual wireless transmission, implementing these systems in the cloud poses issues such as high latency, response times, and higher energy consumption in wearable devices. On the other hand, deploying models entirely on edge wearable sensors faces difficulties due to prohibitive computational costs, like many floating-point operations (FLOPs), associated with deep learning (DL) methodologies. Model compression techniques, such as pruning~\cite{xiaolin2021pruning}, have been utilized to address complexity challenges. Additionally, lightweight machine learning (ML) models for edge devices have been investigated~\cite{Raghu20182cate,malakouti2023heart}. However, fully edge-native approaches often sacrifice system accuracy due to quantization losses or other trade-offs, they do not fully resolve the trade-offs between energy efficiency, latency, and accuracy in continuous ECG monitoring systems. The challenges associated with centralized methods for biosignal inferencing highlight the need for distributed frameworks to overcome these limitations. This study aims to answer how decentralized processing frameworks can improve the efficiency, accuracy, and energy consumption of ECG signal classification in wearable devices. We focus on optimizing DL model configurations with EEPs to reduce data transmission and processing complexity while maintaining high classification performance.

DCenNet offers significant advantages in biosignal classification, particularly through its ability to process data locally. This reduces inferencing latency and energy consumption, key factors for wearable devices and continuous monitoring. The flexibility of these systems allows for easy scalability by adding nodes or devices as needed, enabling seamless adaptation to increasing processing demands. Importantly, their decentralized nature enhances system resilience, as operations can continue even in the event of node failure, reducing the risk of downtime or data loss. Rather than being limited by a single type of signal, this framework can be adapted to various biosignal types like EEG or EMG with appropriate neural network models, making it versatile and cost-effective. By minimizing reliance on expensive infrastructure and reducing data transmission, decentralized systems offer a compelling alternative to traditional centralized approaches.

In this paper, we propose DCenNet, a novel decentralized multistage inferencing approach shown in Fig.~\ref{fig:diagram}. The first approach involves incorporating a single EEP into the network architecture, allowing the large model to be partitioned into two parts and deployed on different nodes. As a second step, we extend it further by introducing two EEPs, dividing the large model into three parts and deploying them on three separate nodes. By leveraging EEPs, a single CNN model can be partitioned into multiple sub-networks~\cite{branchynet}. Each sub-network can be deployed on separate nodes, providing highly confident classifications. The model's cumulative performance, which includes all sub-networks, remains high and comparable to that of the large baseline model. The initial sub-network has lower complexity, allowing it to fit into resource-constrained edge devices better. Because the majority of inputs can be accurately classified by the initial sub-networks, their processing can be terminated early without traversing the remaining network, resulting in lower energy consumption, saved network bandwidth, and improved overall system efficiency, including lower latency and complexity.

Our proposed DCenNet, which employs models with either one or two EEPs, holds significant potential to advance wearable biomedical sensing technology. These approaches greatly lessen reliance on communication networks, reduce latency, and enhance energy efficiency. For instance, utilizing one EEP results in an impressive 94.54\% reduction in wireless data transmission and a 21\% decrease in system complexity while maintaining stable classification accuracy and sensitivity. When two EEPs are utilized, the system achieves a sensitivity of 98.36\% and accuracy of 97.74\%, along with further reductions in wireless data transmission by 91.86\% and complexity by 22\%.

The main contributions of this work are listed below:
\begin{itemize}

 \item DCenNet, a distributed and decentralized multistage inferencing architecture for anomaly detection of biomedical signals is proposed. 
 \item Partitioning a large CNN network into multiple sub-networks with EEPs to deploy the model across different nodes in the Edge-Cloud continuum.
 \item Exploration of various EEP combinations for the proposed architecture and finding a trade-off between complexity and performance utilizing a genetic algorithm.
 \item The proposed approach addresses the limitations associated with a centralized cloud-based system in terms of high inferencing latency and energy consumption, as well as an edge-based system in terms of limited performance and computational capability.
 
\end{itemize} 

Section~\ref{related_work} illustrates some background related to previous work on biomedical signals classification and decentralized approaches. We introduce the baseline architecture for inferencing and proposed decentralized multistage approaches for biomedical signal classification in Section~\ref{model}. Section~\ref{dataset} introduces the dataset used and the methods applied for data preprocessing; the obtained evaluation results on the ECG dataset are in Section~\ref{results}. Section~\ref{conclusion} concludes our work on the proposed distributed multistage decentralized inferencing systems using EEPs. 

%% file: Sections/2_relatedwork.tex
\section{Related Work}
\label{related_work}

The classification of biomedical signals has been extensively studied in the literature. With the emergence of wearable devices, ML models utilizing feature extraction have been applied to various classification tasks, achieving acceptable performance. However, these methods often require substantial computational power, typically centralized on large cloud servers, leading to challenges such as high latency and energy consumption during data transmission~\cite{lu2023end,li2023deep,yan2021nus,sensors_Intelligent}.
It is crucial to explore resource-efficient alternatives that can effectively address these limitations and enable the model to be applicable to a wider range of devices. Thus, deploying small-scale models on wearable devices for classification is proposed to enable rapid inferencing. However, a trade-off arises as the reduced model size may compromise the classifier's performance, leading to the possibility of missing critical signal alerts~\cite{SIVA_TBCas, venkatesan2018knn2categories,plawiak2018novel,sensors_NearSensor}.

Advancements in DL have introduced various techniques to accelerate neural network inferencing, 
model quantization, weight pruning, and model distillation~\cite{xiaolin2021pruning, xiaolin2022ebm, handeepcompress, liu2017learning}. Employing these techniques makes it possible to deploy large models on wearable devices for classification, thereby reducing latency and energy consumption. Despite successfully reducing model size and computational complexity, these methods often involve performance trade-offs.
\cite{handeepcompress} proposed a three-stage pipeline that reduces neural network storage requirements without compromising accuracy. It involves pruning, quantization, and Huffman coding, enabling deployment on embedded systems and mobile apps with improved speedup and energy efficiency. 

\begin{table*}[ht]
\caption{Existing decentralized methods in literature.}
\centering
\setlength{\tabcolsep}{3pt}
\resizebox{\linewidth}{!}{
\begin{tabular}{@{}lcclcccc@{}}
\toprule
\textbf{Authors} & \textbf{Network} &\textbf{Dataset}&\textbf{Methods} & \textbf{Performace}\\ \midrule
Liang~\emph{et~al.}~\cite{liang2022dispense}& VGG19 \& ResNet18 & Cifar10 \& Cifar100 &Dispense Mode&  20\%-50\% Speed$\uparrow$ \text{, Flexible Accuracy} \\
Kang~\emph{et~al.}~\cite{kang2017neurosurgeon}& AlexNet, VGG, MNIST, etc &Speech \& NLP  &Neurosurgeon, DNN Partitioning&3.1$\times$ \text{Latency}$\uparrow$ \text{, 59.5\% Energy}$\downarrow$\\
Yuyang~\emph{et~al.}~\cite{li2022energy}&CNN & Cifar10  &Multi-exit  &43.9\% Energy$\downarrow$ \text{, 0.7\% Accuracy}$\downarrow$ \\
En~\emph{et~al.}~\cite{li2019edge}& DNN &Cifar10  &Edgent, DNN Partitioning, BranchyNet  &Low-latency Edge Intelligence  \\
Pacheco~\emph{et~al.}~\cite{pacheco2020inference}&AlexNet &Cat-and-dog  &BranchyNet Graph, DNN Partitioning & Accelerate DNN Inferencing  \\
       \bottomrule
\end{tabular}
}
\label{tbl:literature_decentralized}
\end{table*}

Table~\ref{tbl:literature_decentralized}
Previous research has proposed distributed learning, where model training and decision-making are distributed across multiple nodes, which can be either end-user devices or servers, and also explored deploying different models or compressed smaller models on various nodes, all of which have been validated in signal classification tasks~\cite{disabato2021distributed,duan2021computation,li2018distributeNetwork}. Multistage decentralized classification has been investigated to increase the effectiveness and accuracy of complex classification tasks.
\cite{li2018distributeNetwork} explored the benefits of Cloud-Edge collaborative inferencing with quantization for Deep Neural Networks (DNN) in mobile intelligent applications. They proposed an auto-tuning neural network quantization framework that analyzes DNN layer characteristics using the ImageNet dataset to enable efficient collaborative inferencing, demonstrating that their framework achieves reasonable network partitions, reduces storage on mobile devices, and maintains high accuracy levels.

BranchyNet has gained considerable interest as a means to optimize DNN inferencing~\cite{branchynet}. It builds upon the concept of network branching to reduce inferencing latency and energy consumption, allowing for early predictions. The concept of EEPs has also emerged as a technique to optimize the inferencing process of DL models. By incorporating EEPs, the model can make intermediate predictions at different stages of the network, allowing for early termination of the inference process when a confident decision is reached. This can significantly reduce computation time, making it particularly beneficial for resource-constrained applications, which contributes to improving the robustness and stability of distributed systems. However, it may result in inconsistent prediction accuracy. Several works have focused on addressing the heavy computational burden caused by DNN on IoT devices and improving computational efficiency~\cite{liang2022dispense,chiang2021optimal,kang2017neurosurgeon,li2022energy,pacheco2020inference,emmons2019cracking}.
\cite{liang2022dispense} proposed a dispense mode to speed up DNN inferencing on IoT devices by efficiently determining the sample's exit position based on difficulty, achieving faster speeds and flexible accuracy adjustments compared to BranchyNet's cascade mode. The performance may not apply to all datasets due to their reliance on sample difficulty. \cite{chiang2021optimal} established the BranchyNet branch placement problem to maximize accurate predictions within a budget of inferencing time. The method uses exit accuracy to determine branch locations, rejecting layers with exit accuracy more than 5\% lower than the original model. Although the algorithm effectively finds optimal branch combinations, it oversimplifies real-world complexity and is sensitive to model inaccuracies.
\cite{kang2017neurosurgeon} proposed Neurosurgeon, a lightweight scheduler that automatically partitions DNN computation at the layer level, achieving significant improvements in latency and energy consumption compared to cloud-only processing. However, it's highly dependent on the neural network structure and hardware, limiting its use on different edge devices.

%% file: Sections/3_model.tex
\section{DCenNet: Decentralized multistage biomedical signal inferencing}
\label{model}

Cloud-native model deployments are well-suited for achieving high model performance. Biomedical signal inferencing typically involves transmitting biomedical data collected by wearable devices to the cloud for analysis and classification. However, this process may encounter several challenges: 1) Firstly, network latencies caused by channel conditions/uncertainties while transmitting a large volume of data to the cloud could lead to delays in inferencing and may be unsuitable for use cases with quick turnaround requirements. 2) Secondly, the large volume of biomedical data from multiple users necessitates high bandwidth for transmission, leading to network congestion and decreased data transfer rates. 3) Finally, the transmission of all collected biomedical data for cloud-only classification and analysis results in a substantial energy expenditure at the edge sensor. Such high-energy consumption can reduce the battery life of wearable devices, thereby diminishing their overall usability. Conversely, edge-native model deployments prioritize low latency and reduced power consumption but may encounter obstacles related to lower model performance and resource constraints. These challenges encompass aspects like model performance, communication bandwidth, sensor power consumption, and the availability of computational and storage resources.

\begin{figure*}[ht]
    \centering
    \includegraphics[width=\linewidth]{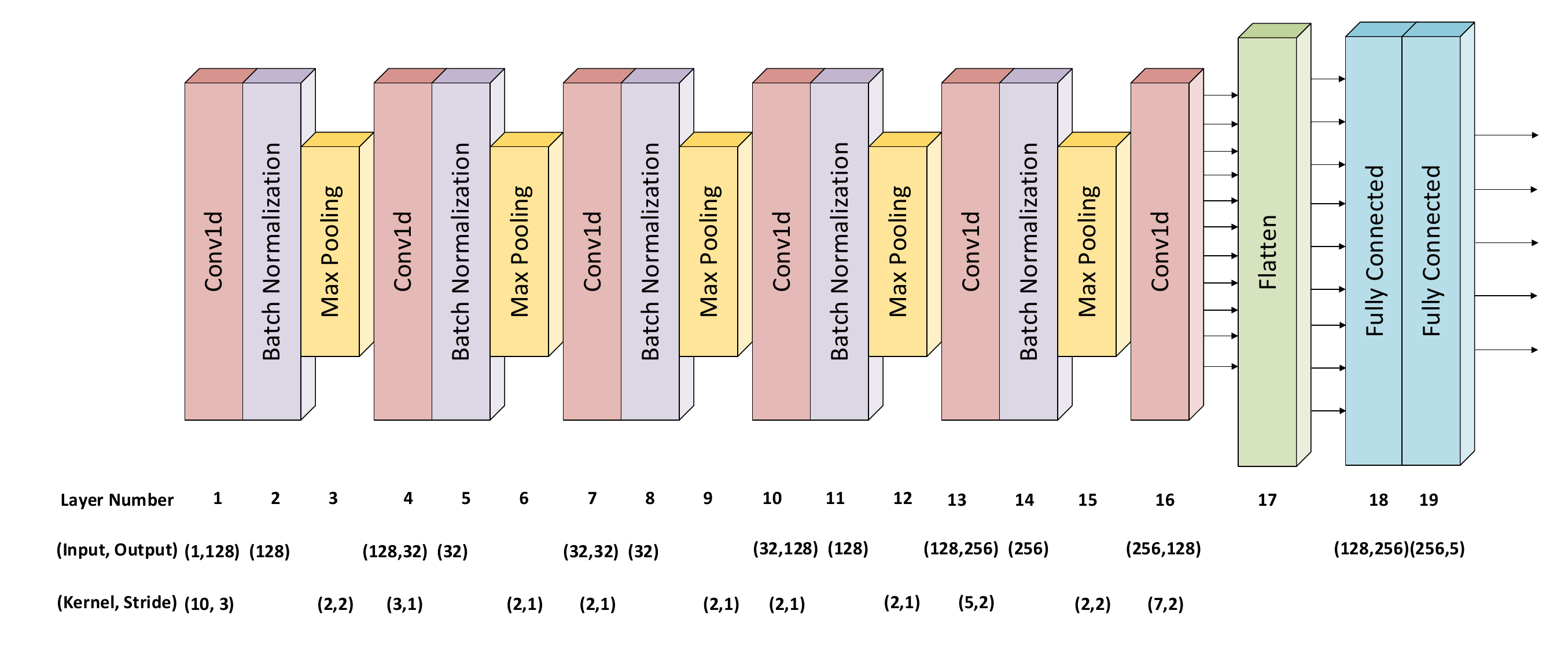}
    \caption{Huge CNN architecture for biomedical signal inferencing.}
    \label{fig:hugeCNN}
\end{figure*}
Deploying DL inferencing processes on a single node (say edge, fog, or cloud) comes with a set of difficulties. To achieve the best of all worlds and address these challenges, we propose a novel approach — DCenNet, a decentralized DL multistage inferencing model that spans multiple nodes across either the Edge-Cloud or the Edge-Fog-Cloud continuum as shown in Fig.~\ref{fig:diagram}. This multistage classifier leverages a cascaded ensemble technique, originally proposed in~\cite{CascadedEn1, CascadedEn2}, to generate partial results at each processing stage if the result is reliable allowing for a timely response to abnormal signals while also mitigating the impact of network failures by distributing computation and results across nodes. These partial results can be utilized independently and incrementally aggregated with previous results, leading to improved cumulative performance. The initial sub-network can better fit onto an edge device with limited resources because it has fewer layers (lower complexity). The processing of normal signals can be stopped early without going through the entire network because the majority of them are classified by the initial sub-networks. This lowers the complexity and latency of inferencing while saving network bandwidth and the energy used by the edge sensor. We test the proposed DCenNet for biomedical signal inferencing by building upon a larger CNN model as illustrated in Fig.~\ref{fig:hugeCNN}. 

A major challenge in deploying decentralized neural networks across multiple nodes is the need to transmit large feature maps between sub-networks, which can lead to significant energy consumption, especially for IoT wearable devices. To mitigate this, we propose integrating an encoder-decoder pair at each EEP to reduce the size of the data transmitted between nodes. The encoder compresses the feature maps before transmission, significantly lowering the energy required for wireless communication. Once the data reaches the next sub-network, the decoder reconstructs the feature maps for further processing. This approach ensures that only the most essential information is transmitted, making the system more energy-efficient. The advantage over traditional centralized methods is that instead of sending all inputs directly to the cloud without discrimination, only the data that cannot be classified at earlier stages is transmitted, as described in Algorithm~\ref{alg:stage1}. This selective transmission reduces network load, extends device battery life, and facilitates more efficient processing across edge devices.

Given the vulnerability of IoT devices to security threats, safeguarding data and ensuring system reliability are critical concerns. In our proposed decentralized approach, the distribution of model computations across multiple nodes enhances resilience against potential network interruptions. By processing data locally on resource-constrained edge devices, our system reduces reliance on cloud infrastructure, minimizing the transmission of sensitive data and thus mitigating risks associated with data breaches or interception during communication. This local processing approach helps protect both data privacy and system security. Additionally, the decentralized framework distributes the workload across different nodes, improving system efficiency and reducing the impact of potential security threats.

\begin{algorithm}[h]
\caption{ECG Signal Processing and Model Evaluation}
\label{alg:stage1}
\begin{algorithmic}[1]
\REQUIRE Raw ECG signal data
\ENSURE Performance and Complexity Comparison
\STATE \textbf{Model Input:} Feature matrix for all ECG beats
\STATE \textbf{Model Output:} Probability
\STATE \textbf{Confidence threshold:} 0 \TO 1 with step $0.01$
\STATE Data preprocessing and feature extraction
\FOR{each ECG beat}
    \STATE Calculate $pred$ for each class in each EEP of the model
    \STATE Select the class with the highest probability as the outcome
\ENDFOR

\FOR{$\text{threshold} = 0$ \TO $1$}
    \STATE Initialize variables for performance and complexity
    \IF{$pred$ $> \text{threshold}$}
        \STATE Exit early
    \ELSE
        \STATE Transmit to the next stage
        \FOR{$\text{threshold} = 0$ \TO $1$}
            \IF{$pred$ $> \text{threshold}$}
                \STATE Exit early
            \ELSE
                \STATE Transmit to the next stage
                \STATE Exit at the original point
            \ENDIF
        \ENDFOR
    \ENDIF
    
    \STATE Calculate performance 
    \STATE Calculate complexity 
    \STATE Record performance and complexity for the current threshold
\ENDFOR
\end{algorithmic}
\end{algorithm}

\subsection{Single EEP}
\begin{figure}[!htb]
    \centering
    \includegraphics[width=\linewidth]{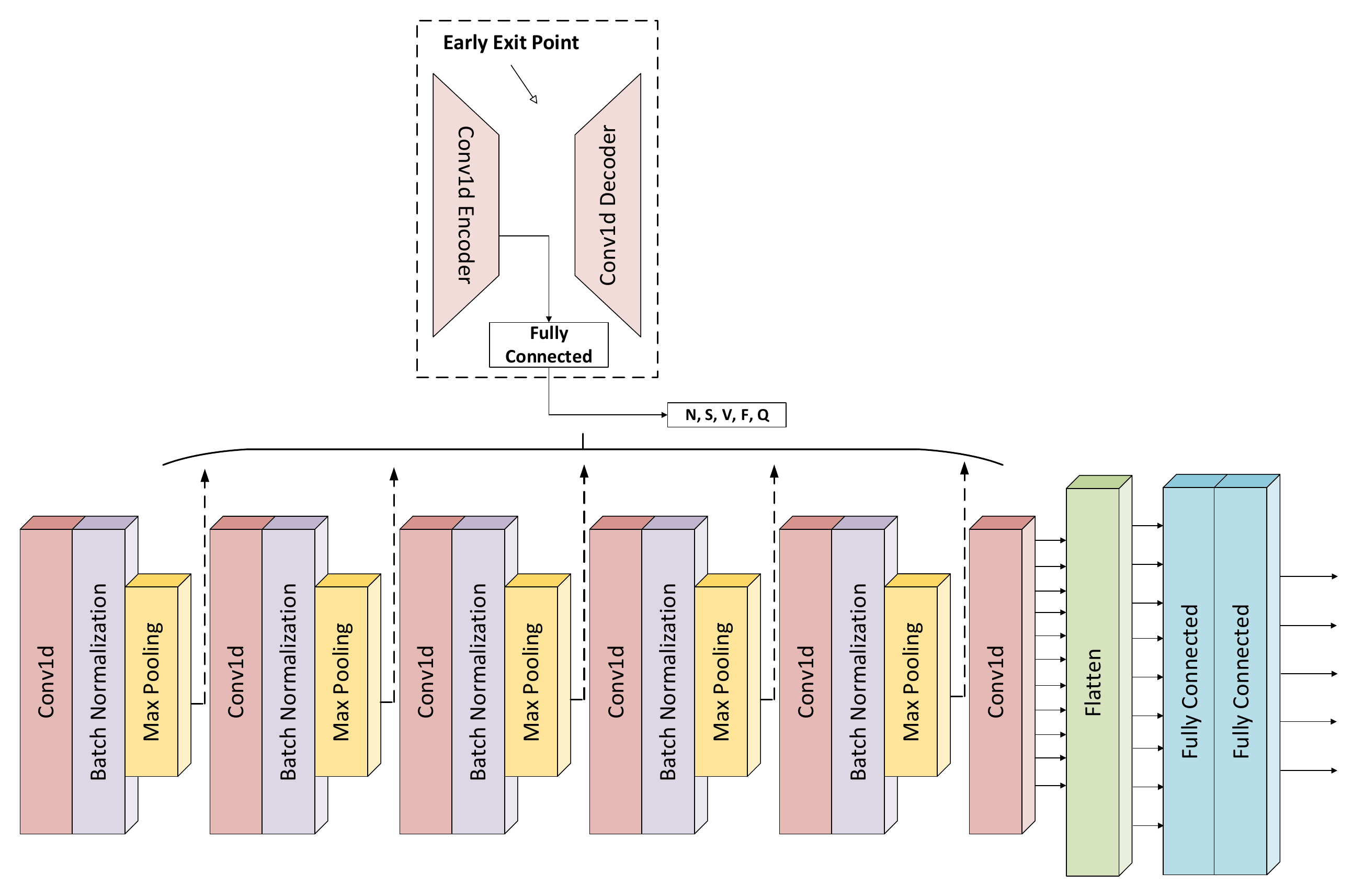}
    \caption{The architecture of the Edge-Cloud continuum for decentralized multistage biomedical signal inferencing using single EEP.}
    \label{fig:singleExitPointArchitecture}
\end{figure}

For the Edge-Cloud continuum, we extend the network architecture by incorporating a single additional EEP, as illustrated in Fig.~\ref{fig:singleExitPointArchitecture}. The EEP can correspond to any one of the directions indicated by each individual dashed arrow shown in Fig.~\ref{fig:singleExitPointArchitecture}. By placing the EEP within the network, we aim to achieve a balance between achieving optimal performance and managing computational complexity demands between edge-native and cloud-native deployments. To achieve this, a series of experiments were conducted to investigate the impact of different EEP positions on the overall system performance and complexity. Our objective is to identify the best location for the EEP regarding different targets, considering the trade-off between the system's inferencing performance and its computational resources. By analysing the experimental outcomes, we could determine the best-suited EEP location, that aligns with the requirements and constraints of the Edge-Cloud continuum. 
The placement of the EEP effectively optimized the model's inferencing process, providing low inferencing latency and efficient resource utilization while maintaining high accuracy in biosignal classification tasks. Assuming the CNN consists of $L$ convolutional layers, in the Edge-Cloud continuum, which incorporates a single EEP, there can be $L-1$ potential positions for the EEP. Table~\ref{tbl:singleExitPoint} displays all the possibilities of convolutional layers in edge and cloud layouts when incorporating a single EEP only. For a network comprising six convolutional layers, there are five potential positions for EEP. 

\begin{table}[!htb]
\centering
\caption{Possibilities of convolutional layer deployment on the Edge-Cloud continuum with a single EEP.}
\resizebox{0.8\linewidth}{!}{
\begin{tabular}{@{}cccccc@{}}
\toprule
                                    & \multicolumn{5}{c}{\textbf{Deployment Positions for Convolutional Layers}}                                                                             \\ \midrule
\multicolumn{1}{c|}{\textbf{Edge}}  & \multicolumn{1}{c|}{1}         & \multicolumn{1}{c|}{1,2}     & \multicolumn{1}{c|}{1,2,3} & \multicolumn{1}{c|}{1,2,3,4} & \multicolumn{1}{c|}{1,2,3,4,5} \\ \cmidrule(l){2-6} 
\multicolumn{1}{c|}{\textbf{Cloud}} & \multicolumn{1}{c|}{2,3,4,5,6} & \multicolumn{1}{c|}{3,4,5,6} & \multicolumn{1}{c|}{4,5,6} & \multicolumn{1}{c|}{5,6}     & \multicolumn{1}{c|}{6}         \\ \bottomrule
\end{tabular}}
\label{tbl:singleExitPoint}
\end{table}

\subsection{Two EEPs}
\begin{figure}[!htb]
    \centering
    \includegraphics[width=\linewidth]{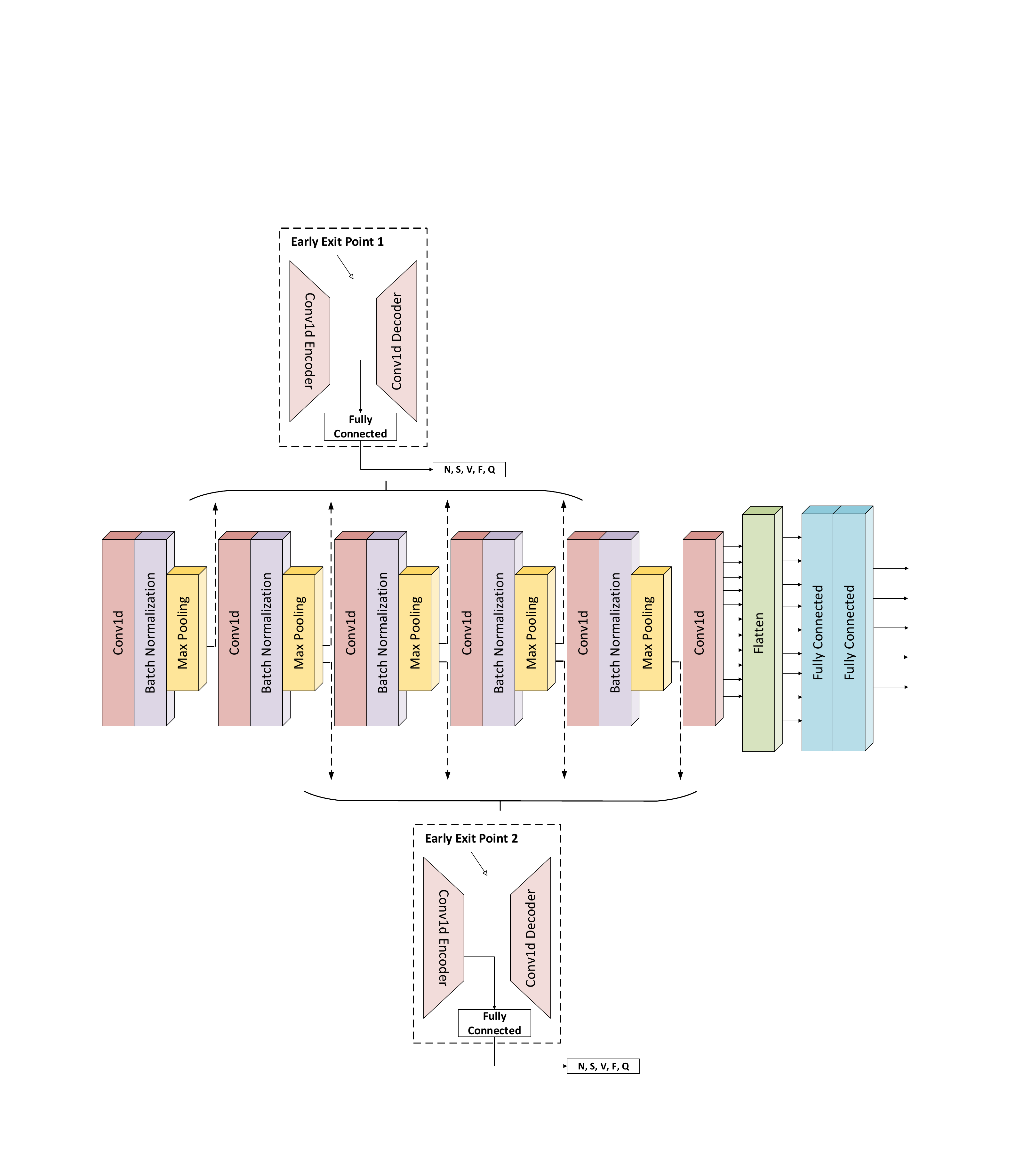}
    \caption{The architecture of the Edge-Fog-Cloud continuum for decentralized multistage biomedical signal inferencing using two EEPs.}
    \label{fig:TwoExitPointsArchitecture}
\end{figure}

For the Edge-Fog-Cloud continuum, we augmented the existing network architecture by introducing two possible EEPs, as illustrated in Fig.~\ref{fig:TwoExitPointsArchitecture}. Two EEPs were incorporated to accommodate the varying demands and characteristics of edge-native, fog-native, and cloud-native deployments. To determine the most effective exit location for these EEPs, we conducted a series of experiments, exploring numerous combinations of EEP placements. Our objective is to identify the optimal locations for the two EEPs, aligning with the specific requirements and constraints imposed by the Edge-Fog-Cloud continuum. The integration of the dual EEPs effectively fine-tuned our decentralized distributed system, making it adaptable and robust in various deployment scenarios across the Edge-Fog-Cloud continuum. This enabled the system to adapt effectively to various environments, guaranteeing low complexity in edge-native settings, optimizing resource utilization in fog-native architectures, and providing high performance in cloud-native configurations. Assuming the CNN consists of $L$ convolutional layers, in the Edge-Fog-Cloud continuum with two EEPs, there exists a range of options, comprising $\sum_{i=1}^{L-2}i$ possibilities. Table~\ref{tbl:twoExitPoint} presents the various scenarios of convolutional layers in Edge-Fog-Cloud continuum when integrating two EEPs. In the case of a six-layer convolutional network, there are a total of 10 potential combinations for two EEPs. 

\begin{table}[!htb]
\centering
\caption{Possibilities of convolutional layer deployment on the Edge-Fog-Cloud continuum with two EEPs.}
\resizebox{\linewidth}{!}{
\begin{tabular}{@{}ccccccccccc@{}}
\toprule
                                    & \multicolumn{10}{c}{\textbf{Deployment Positions for Convolutional Layers}}                                                                                                                                                                                                                 \\ \midrule
\multicolumn{1}{c|}{\textbf{Edge}}  & \multicolumn{4}{c|}{1}                                                                                                & \multicolumn{3}{c|}{1,2}                                                           & \multicolumn{2}{c|}{1,2,3}                          & \multicolumn{1}{c|}{1,2,3,4} \\ \cmidrule(l){2-11} 
\multicolumn{1}{c|}{\textbf{Fog}}   & \multicolumn{1}{c|}{2}       & \multicolumn{1}{c|}{2,3}   & \multicolumn{1}{c|}{2,3,4} & \multicolumn{1}{c|}{2,3,4,5} & \multicolumn{1}{c|}{3}     & \multicolumn{1}{c|}{3,4} & \multicolumn{1}{c|}{3,4,5} & \multicolumn{1}{c|}{4}   & \multicolumn{1}{c|}{4,5} & \multicolumn{1}{c|}{5}       \\ \cmidrule(l){2-11} 
\multicolumn{1}{c|}{\textbf{Cloud}} & \multicolumn{1}{c|}{3,4,5,6} & \multicolumn{1}{c|}{4,5,6} & \multicolumn{1}{c|}{5,6}   & \multicolumn{1}{c|}{6}       & \multicolumn{1}{c|}{4,5,6} & \multicolumn{1}{c|}{5,6} & \multicolumn{1}{c|}{6}     & \multicolumn{1}{c|}{5,6} & \multicolumn{1}{c|}{6}   & \multicolumn{1}{c|}{6}       \\ \bottomrule
\end{tabular}}

\label{tbl:twoExitPoint}
\end{table}

%% file: Sections/3.3_EmbedSys.tex
\subsection{Embedded System Evaluation}

Implementing a model in resource-constrained IoT devices presents several challenges. First, IoT devices typically have limited computational power and memory, which makes it difficult to deploy complex models. The introduction of EEP helps mitigate this by allowing devices to process only smaller portions of the model, thereby reducing the computational load. Second, power consumption is a critical concern, especially for battery-operated devices. The use of EEP significantly lowers computational demands, leading to an average energy savings of 73.6\% as observed in this study. Third, decentralized systems often face latency issues due to data transmission between nodes. However, the early exit strategy reduces the volume of data transmitted, thereby minimizing delays. Finally, bandwidth limitations, which are common in IoT environments, are effectively addressed by reducing the amount of data transmitted through the use of EEPs. 

By incorporating the EEP, the average FLOPs required are significantly reduced. Fewer FLOPs mean that the inferencing will require fewer computational resources at the edge, and also wireless transmission and further processing can be avoided if not necessary. When data isn't transmitted, the edge device can operate in a low-power mode, thereby achieving energy savings. Compared to continuous Bluetooth transmission, our method reduces the average values by 67.3\%, achieving reductions of 73.7\% in broadcast mode and 61.7\% in connected mode. The details are in Table~\ref{tb:power_comparison}.

\begin{table}[!htb]
\centering
\caption{Power Consumption (mA) of Various Methods Across Deployment Thresholds}
\resizebox{0.7\linewidth}{!}{
\begin{tabular}{@{}lccccc@{}}
\toprule
\textbf{Confidence} & \textbf{0.5} & \textbf{0.6} & \textbf{0.7} & \textbf{0.8} & \textbf{0.9} \\ \midrule
Inference Only   & 0.74 & 0.74 & 0.74 & 0.74 & 0.74 \\
Sleep Mode       & 0.58 & 0.58 & 0.58 & 0.58 & 0.58 \\ 
Connected Mode   & 3.66 & 3.66 & 3.66 & 3.66 & 3.66 \\
Broadcast Mode   & 3.20 & 3.20 & 3.20 & 3.20 & 3.20 \\
\textbf{Ours (Connected)} & 1.07 & 1.16 & 1.28 & 1.71 & 1.79 \\
\textbf{Ours (Broadcast)} & 0.79 & 0.80 & 0.82 & 0.86 & 0.94 \\
\bottomrule
\end{tabular}}
\label{tb:power_comparison}
\end{table}

\begin{figure}[!htb]
    \centering
    \includegraphics[width=\linewidth]{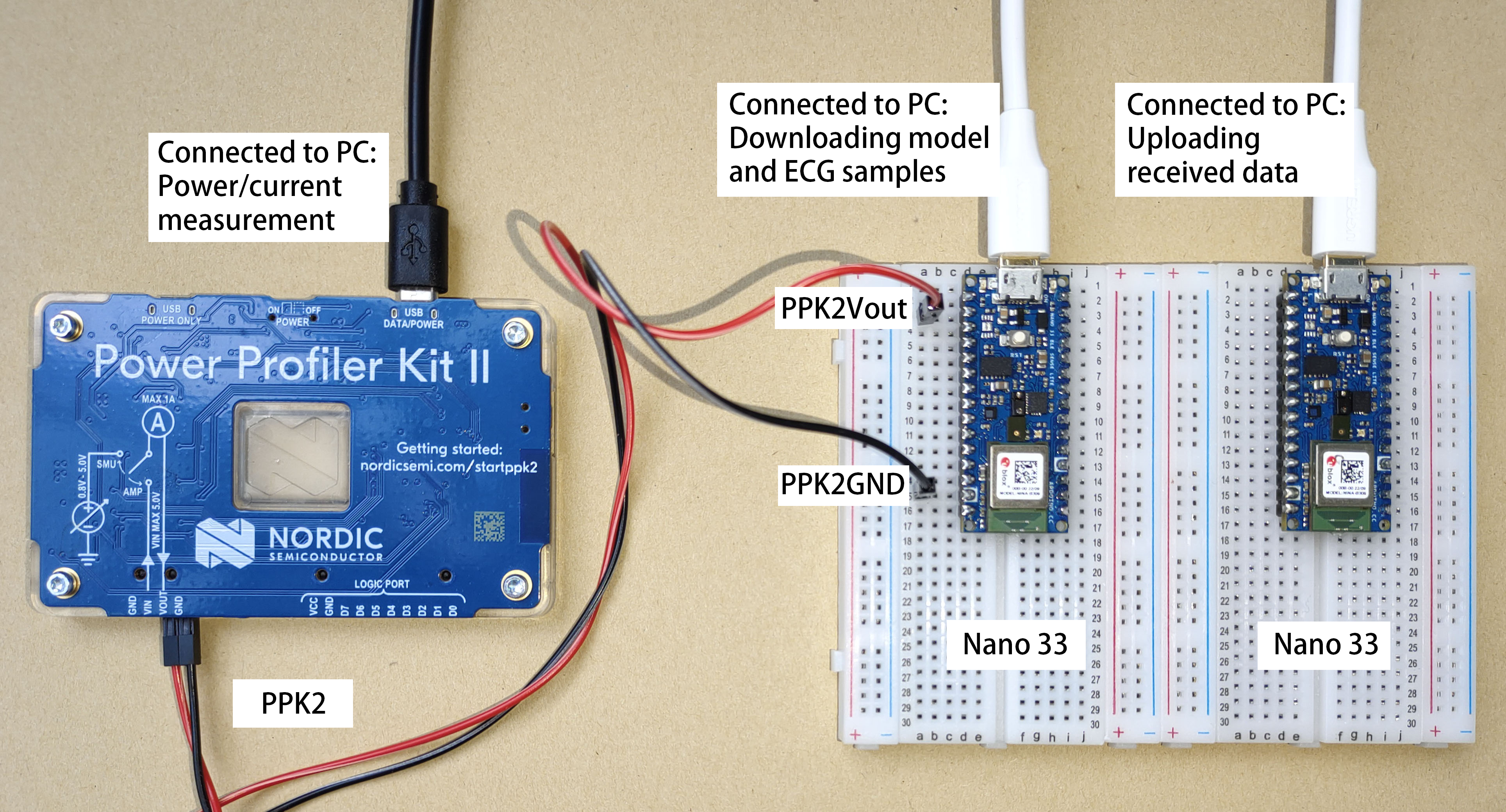}
    \caption{Hardware implementation: PPK2 for measuring power consumption, Nano 33 (left) running the model, and another Nano 33 (right) serving as a receiver for the BLE signal.}
    \label{fig:Hardware}
\end{figure}

We deployed part of our model, i.e. sub-network until the first early exit, on an Arduino Nano 33 BLE Sense embedded platform (Nano 33) to evaluate its power consumption when deployed in an edge device. 
We compiled our model and deployed the main control logic of the firmware, such as data input and output logic, power-saving functions, and BLE transmission logic on the BLE Sense board. 
Fig. \ref{fig:Hardware} illustrates the experimental setup, which includes the Nano 33 and the Nordic Power Profiler Kit II (PPK2) used for measuring current consumption. 
The PPK2 has a measurement error of ±10\% below 50 mA, increasing to ±15\% above 50 mA. 
Another Nano 33, also shown in Fig. \ref{fig:Hardware}, serves to receive BLE signals and display them on a PC. 
ECG from MIT-BIH Arrhythmia records were sent to the Nano 33 board, configuring the processing rate to one heartbeat per operation. 

The Nano 33 has a Cortex-M4 ARM core with BLE integrated. The popularity of Arduino led us to choose it as the implementation platform, making our work easy for other teams to reproduce. The Nano 33 board has a maximum of 256 KB RAM and 1 MB ROM. Our decentralized CNN model is designed to prune its layers, making it suitable for resource-limited edge devices. In our experiments, we used a version of a two-layer model that requires about 70 KB of memory. The "decentralized" approach significantly reduces the model size, as the edge device processes only a small portion of the whole network. This allows the Nano 33 to handle tasks efficiently within its memory limits while maintaining system performance. More detailed material is on our GitHub\footnote{\url{https://github.com/microa/DCentNet}}.

\begin{figure}[!htb]
    \centering
    \includegraphics[width=\linewidth]{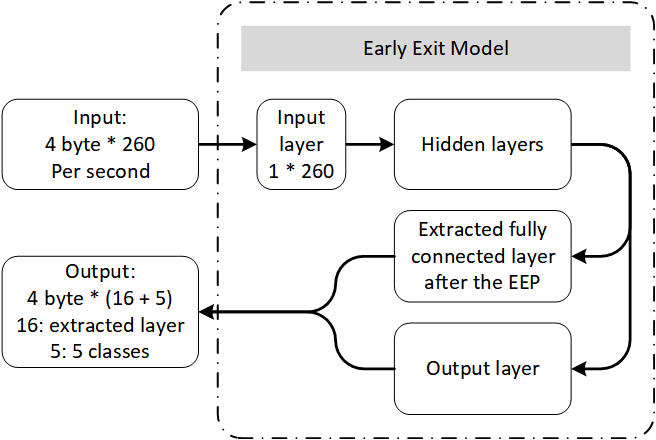}
    \caption{Block diagram illustrating subnetwork on the edge device.}
    \label{fig:Diagram_implementation}
\end{figure}

Fig. \ref{fig:Diagram_implementation} shows the diagram of the model implementation on the reference edge device. 
The input data is 32-bit float-type, with an inference rate nearly equal to a heartbeat per second. 
The original sampling rate of the MIT-BIH dataset is 360Hz and resampled to 260 before being input into the model. 
The model is designed to process one-dimensional data consisting of 260 samples. 
The outputs of the model are divided into two feature groups: the classification results and the fully connected encoder layer. We set thresholds at 0.5, 0.6, 0.7, 0.8, and 0.9 to evaluate performance. Confidence levels below 0.5 were excluded due to their limited reliability, despite potentially lower power consumption.

%% file: Sections/4_dataset.tex
\section{Dataset}
\label{dataset}
An open database of ambulatory ECG records collected from 48 patients under varied circumstances makes up the MIT-BIH Arrhythmia database~\cite{mitbih}. The 48 participants in this dataset, 25 men between the ages of 32 and 89, and 22 women between the ages of 23 and 89, provided ambulatory ECG records. Approximately 60\% of the recordings included in this study were obtained from inpatients. The ECG signals were sampled with an 11-bit resolution and sampled at 360 Hz. Each record is about 30 minutes long.
We evaluated performance using a single ECG lead from the MIT-BIH database, with 70\%, 15\%, and 15\% respectively for training, validation, and testing. The first available lead, which is modified limb lead \uppercase\expandafter{\romannumeral2} (MLII), obtained by placing electrodes across the torso, is used in this experiment. It is determined that 260 samples centred on the R peak provide sufficient details on the signal shape~\cite{260sample}. Based on the AAMI standards, ECG is mapped into 5 classes, i.e. N, SVEB, VEB, F, Q~\cite{AAMI}.

%% file: Sections/5_results.tex
\section{Results \& Discussion}
\label{results}

\begin{figure*}[h]
  \centering
  \subfloat[System accuracy]{\includegraphics[width=0.33\linewidth]{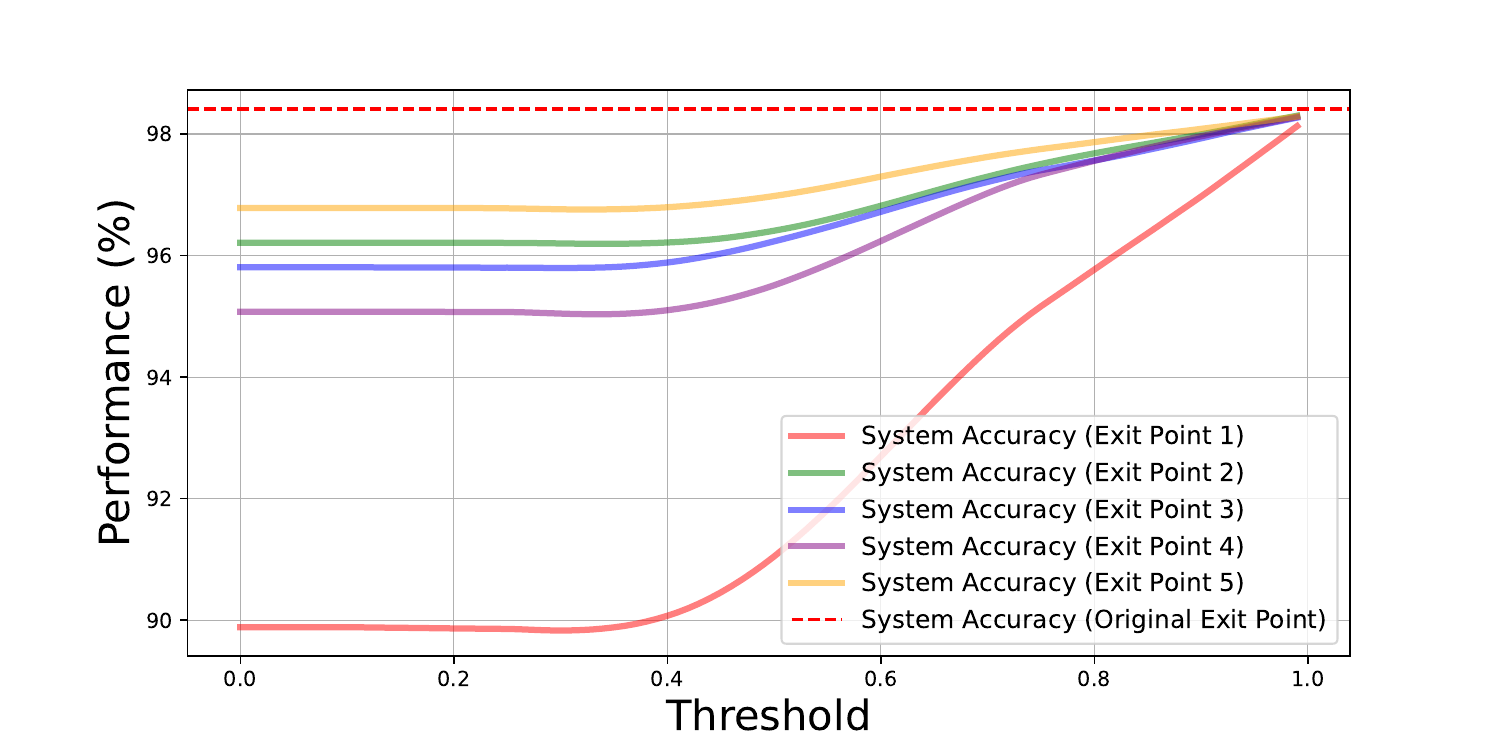}}\hfill
  \subfloat[System sensitivity]{\includegraphics[width=0.33\linewidth]{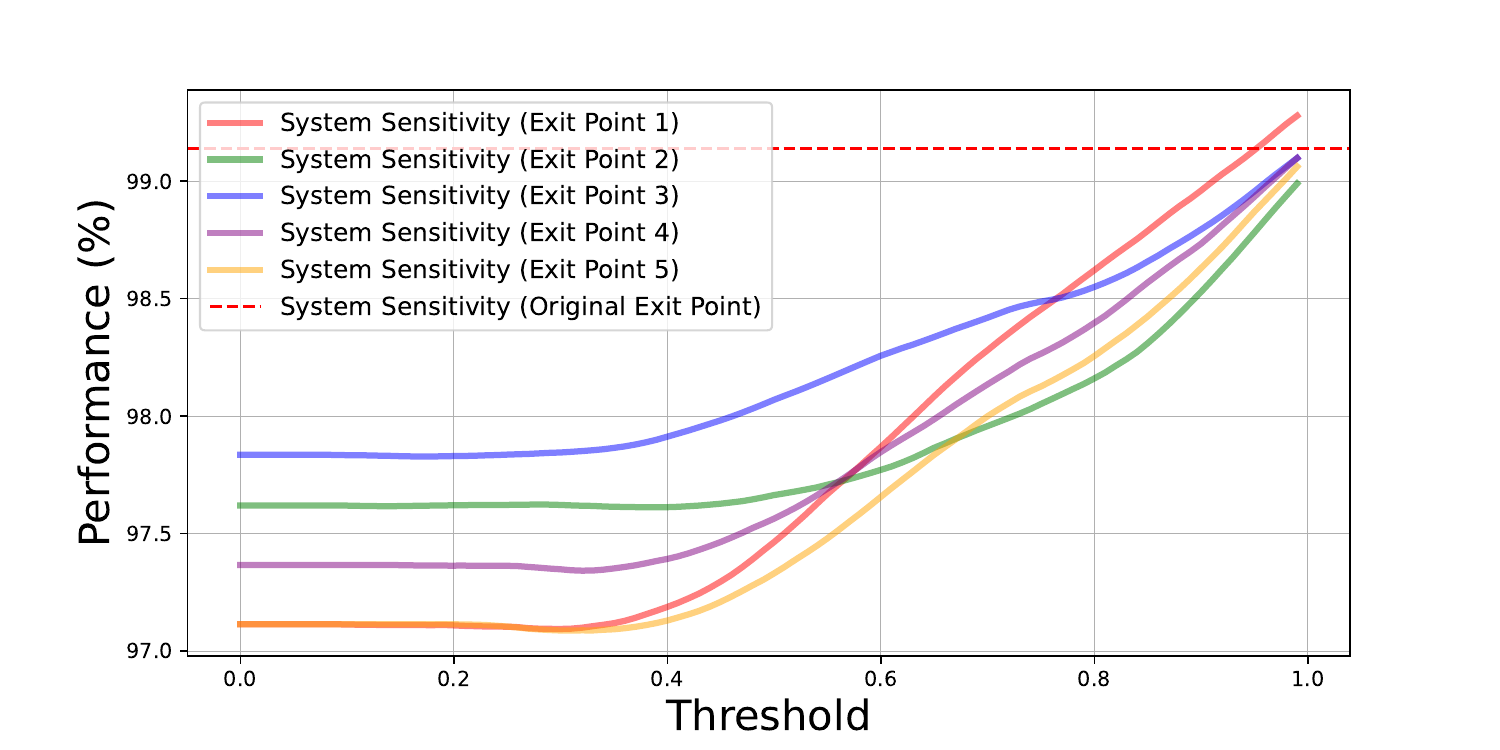}}\hfill
  \subfloat[DtC]{\includegraphics[width=0.33\linewidth]{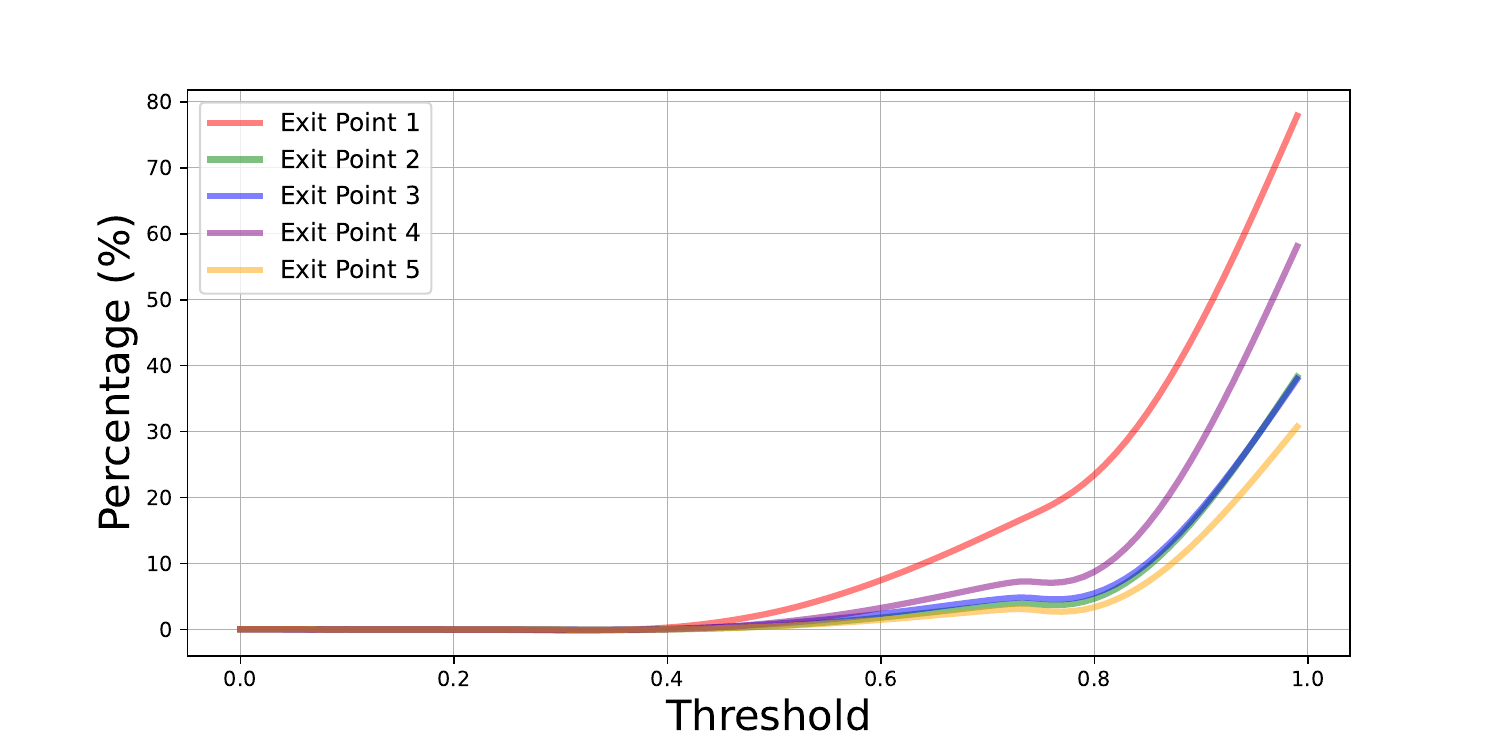}}\hfill
  \subfloat[Total FLOPs]{\includegraphics[width=0.33\linewidth]{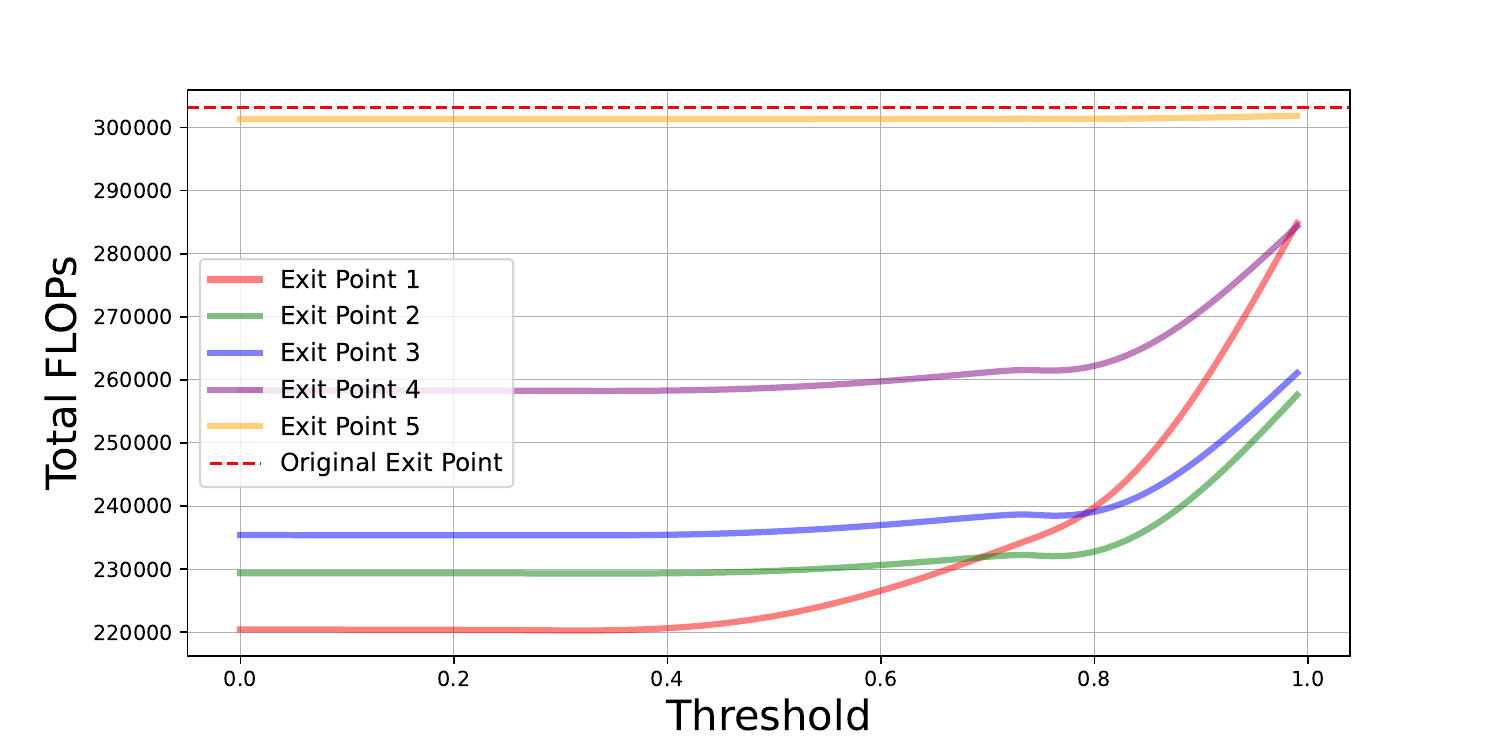}}\hfill
  \subfloat[Exit rate]{\includegraphics[width=0.33\linewidth]{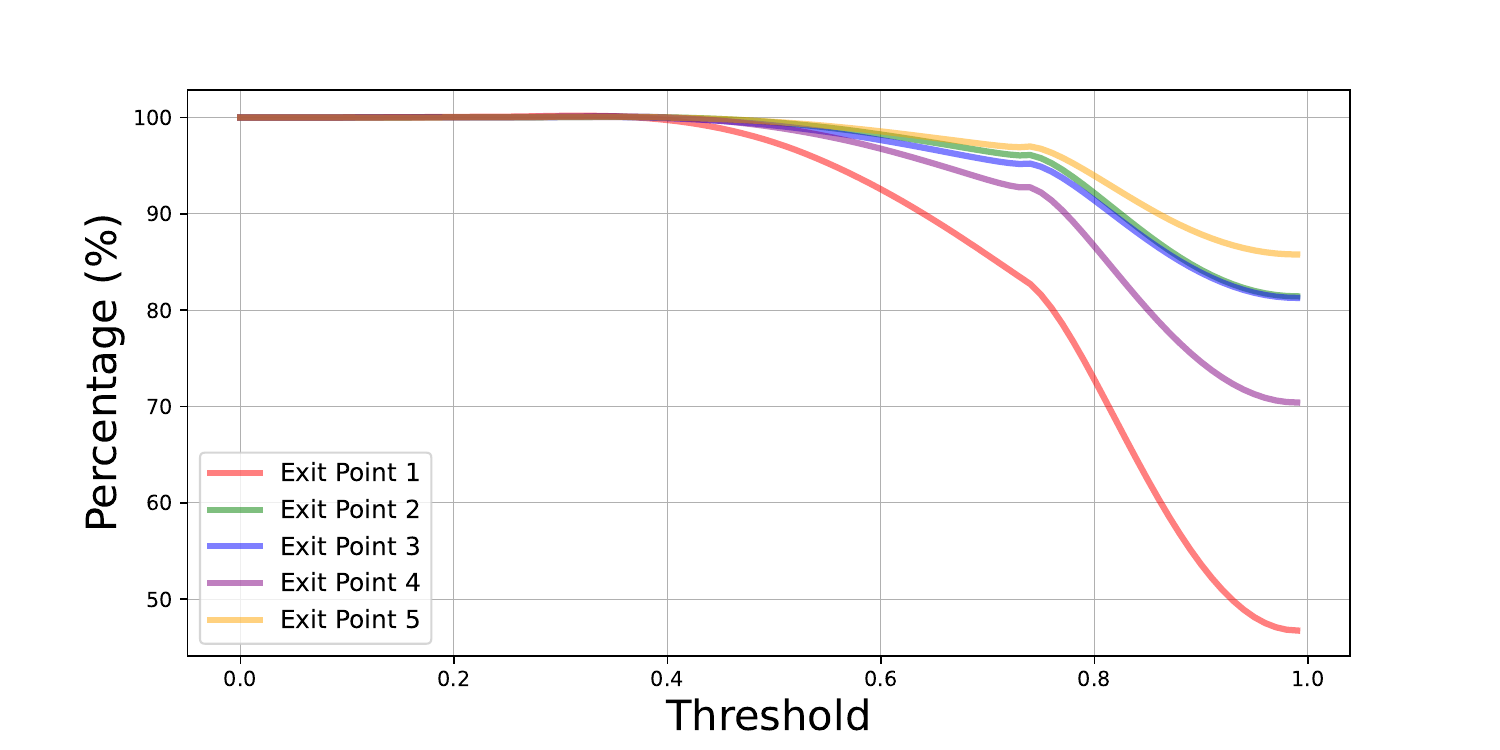}}\hfill
  \subfloat[Efficiency rate]{\includegraphics[width=0.33\linewidth]{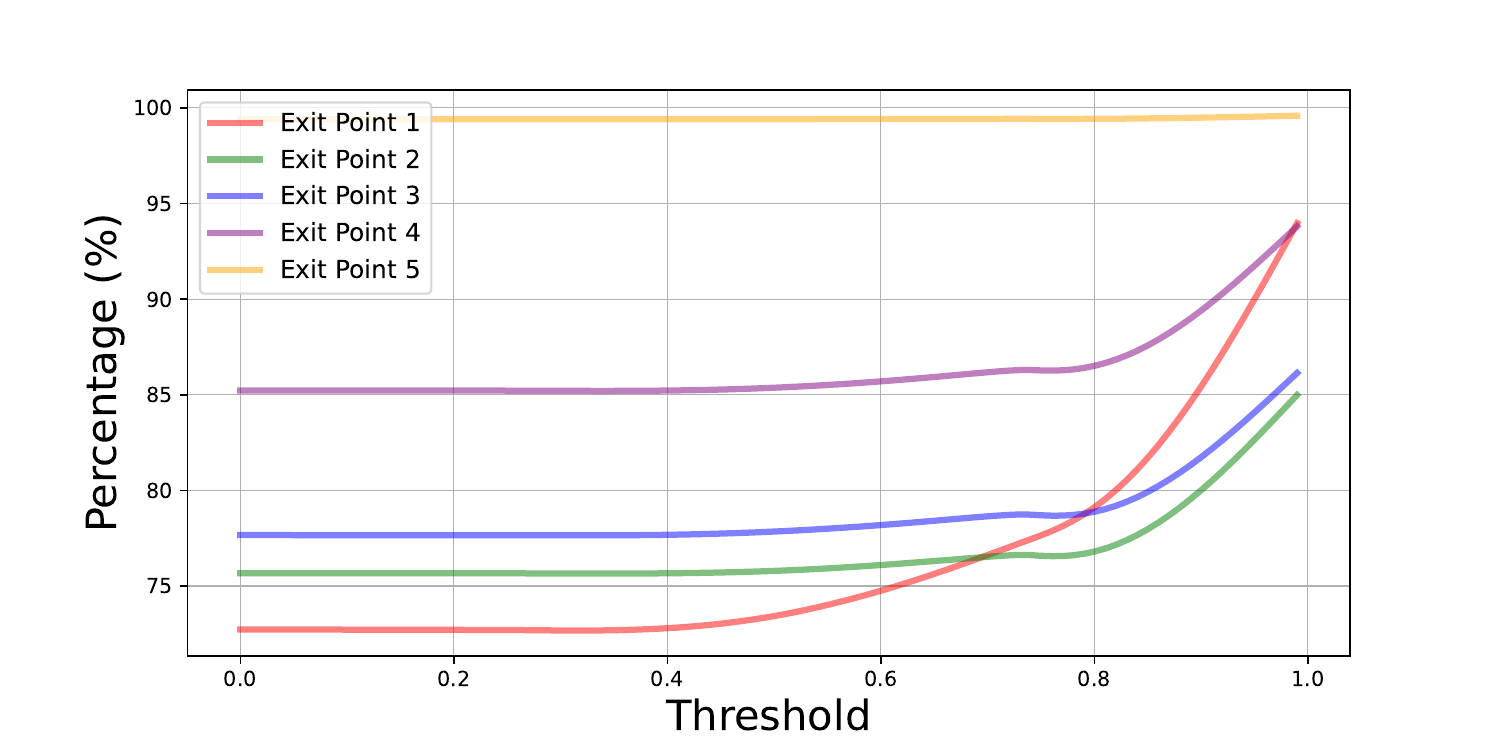}}\hfill
  \caption{Performance, complexity of the model using single EEP.}
  \label{fig:exitPerformance_TwoParts}
\end{figure*}

\begin{figure*}[h]
  \centering
  \subfloat[System accuracy]{\includegraphics[width=0.33\linewidth]{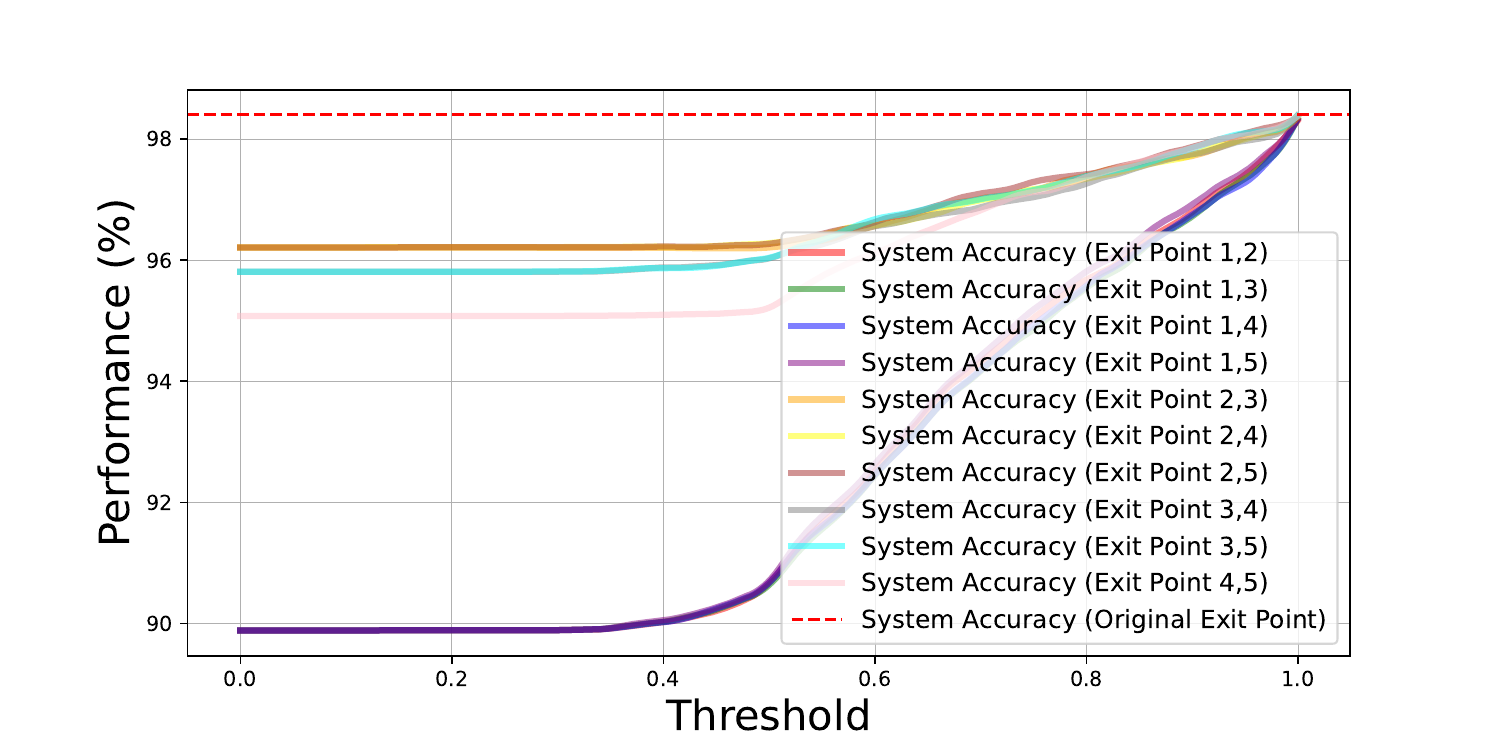}}\hfill
  \subfloat[System sensitivity]{\includegraphics[width=0.33\linewidth]{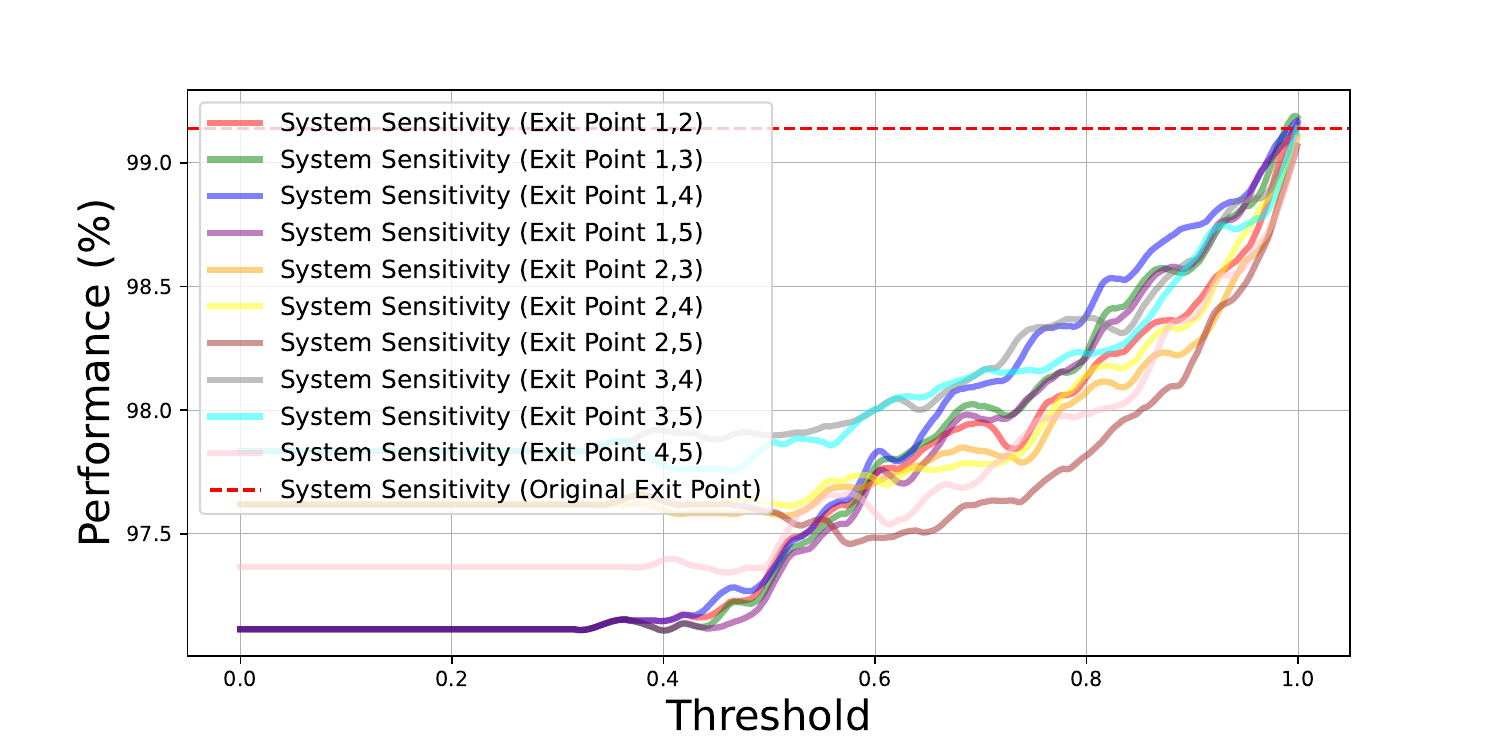}}\hfill
  \subfloat[DtC]{\includegraphics[width=0.33\linewidth]{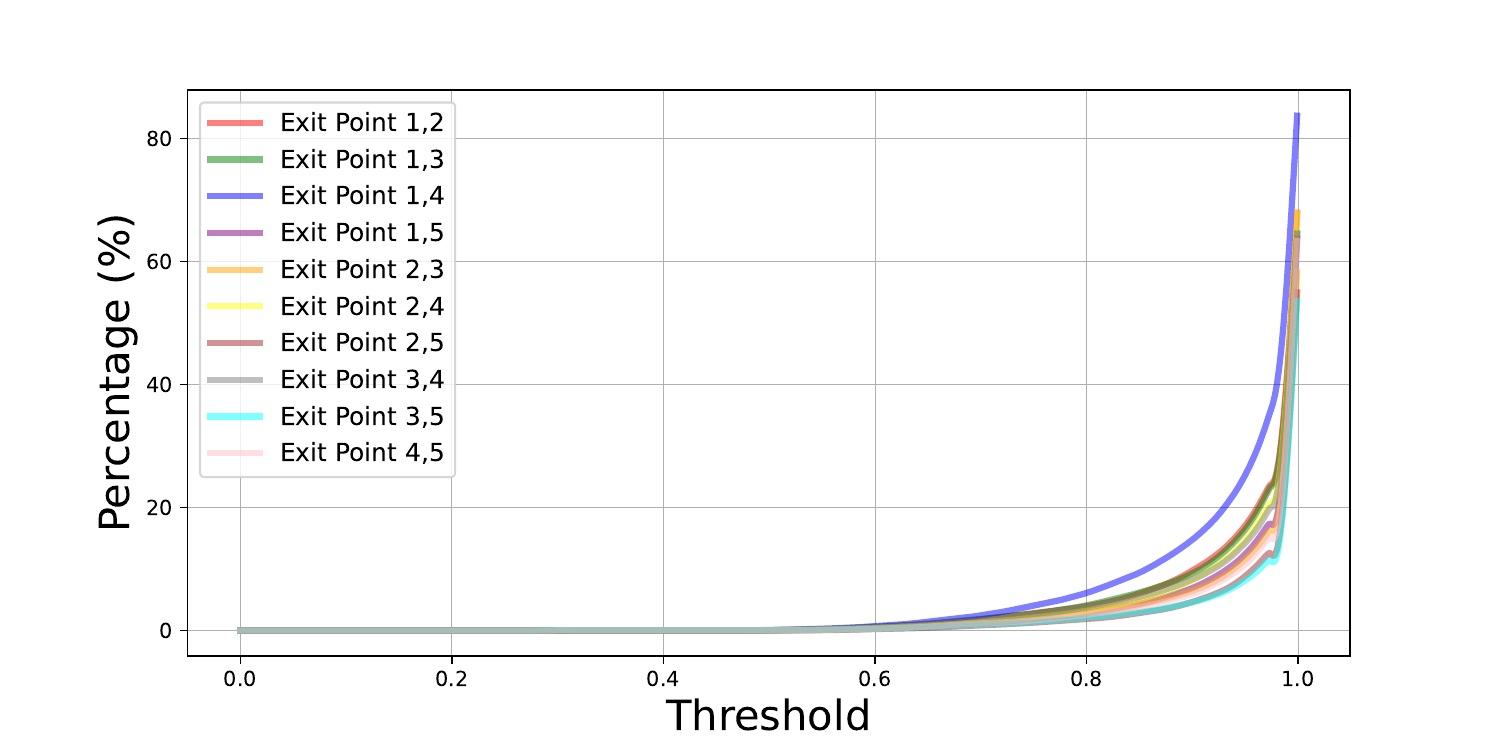}}\hfill
  \subfloat[Total FLOPs]{\includegraphics[width=0.33\linewidth]{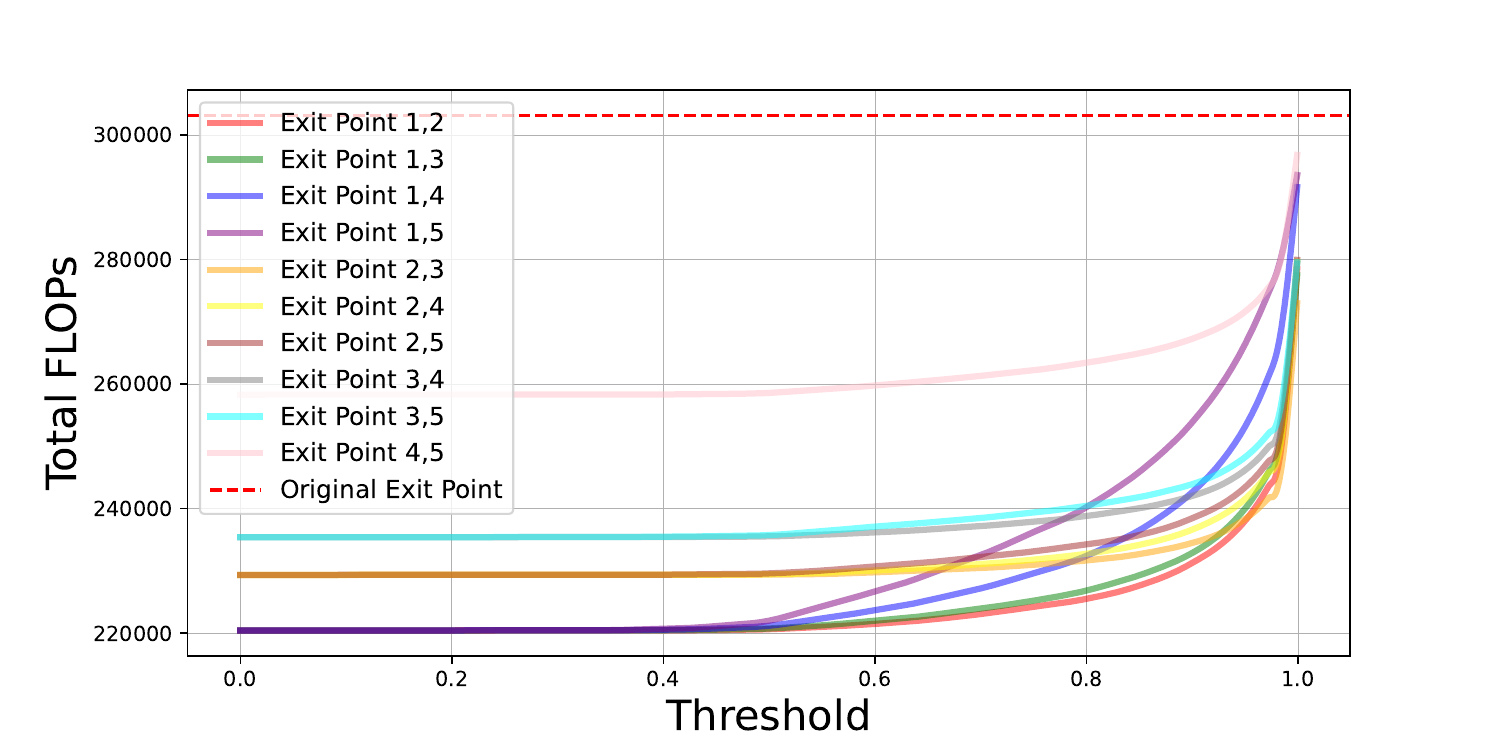}}\hfill
  \subfloat[Exit rate 1]{\includegraphics[width=0.33\linewidth]{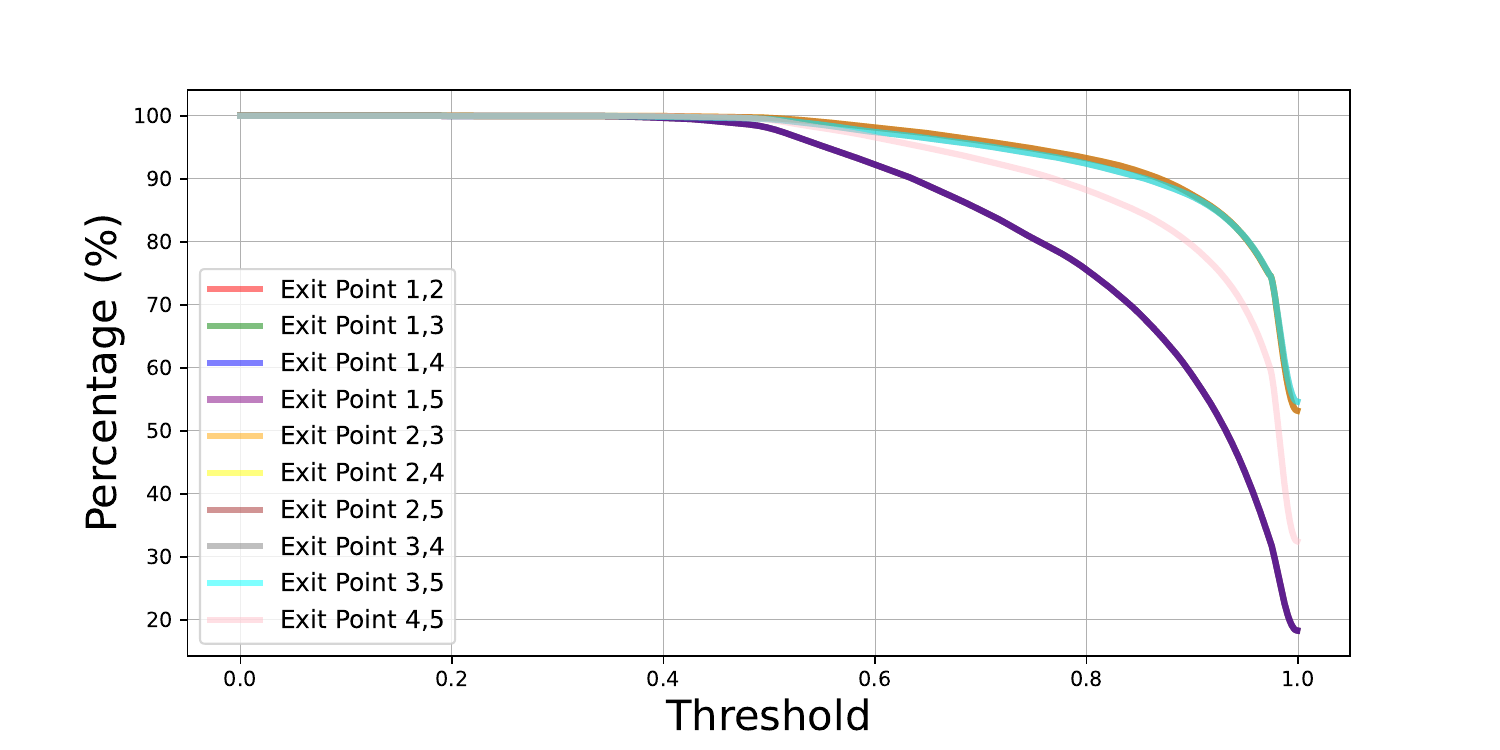}}\hfill
  \subfloat[Exit rate 2]{\includegraphics[width=0.33\linewidth]{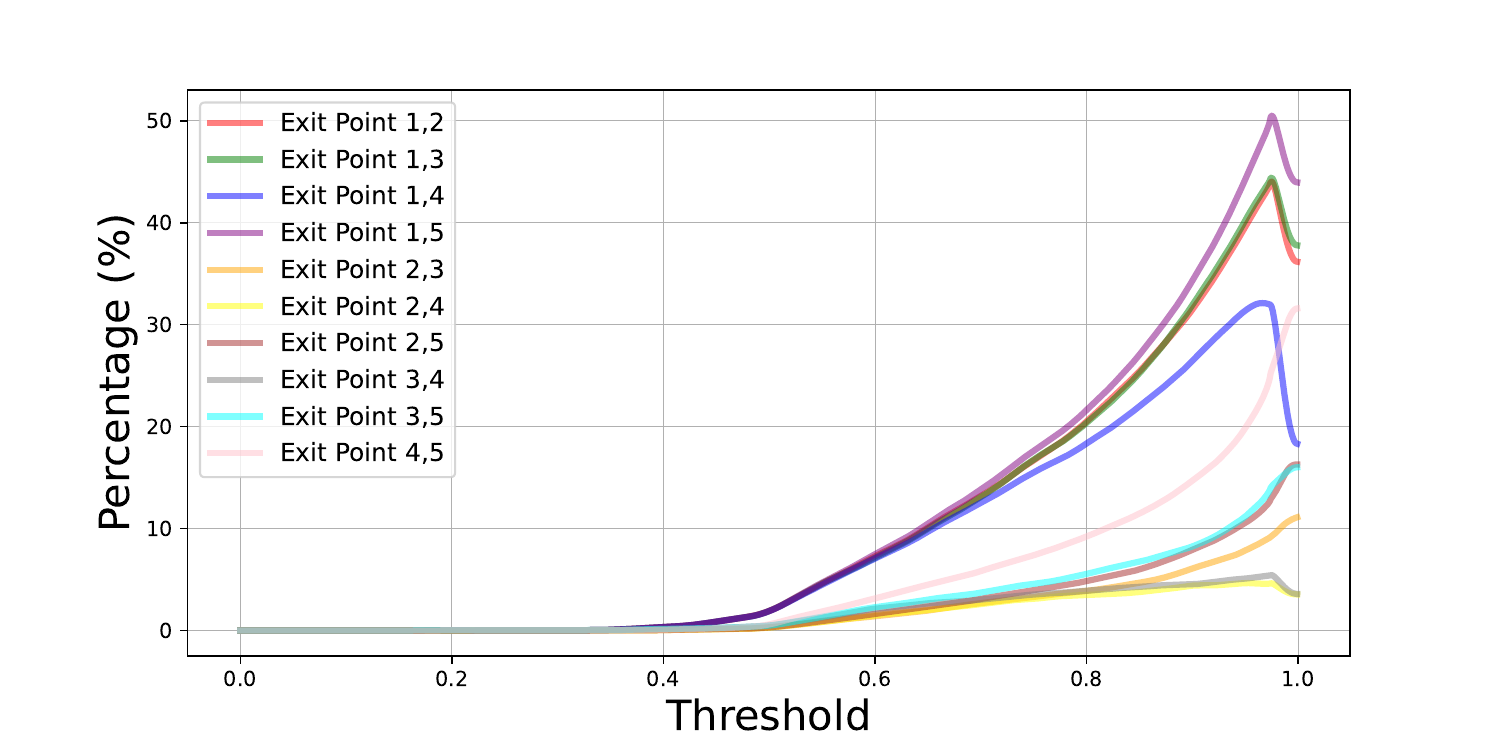}}\hfill
  \caption{Performance, complexity of the model using two EEPs.}
  \label{fig:exitPerformance_ThreeParts}
\end{figure*}

There are numerous possibilities for EEP settings, which increase as the network size grows. Identifying the best early exit point or combinations of early exit points maximises performance and minimizes complexity. The objective is to leverage the experimental findings to identify the best EEP or point combinations that align most closely with the actual requirements. Adjusting the algorithm to meet practical requirements involves assigning weights to performance and complexity based on real-world considerations. Experiments are executed with architectures containing either a single EEP or two EEPs, investigating various positional arrangements. Following this, we employed an algorithm to select the most suitable configuration, which best meets the requirements. The system sensitivity evaluates the percentage of true positive cases accurately identified by the system, while the system accuracy indicates the overall rate of correct classifications relative to all predictions; additionally, DtC represents the percentage of data sent to the cloud out of the total data in the test set~\cite{xiaolin2024two}. The exit rate represents the proportion of outcomes exceeding the confidence threshold, and thereby being deemed reliable relative to the overall count. The efficiency rate denotes the ratio between the total FLOPs of the system after the introduction of the EEP and the FLOPs of the original CNN model.

\begin{algorithm}[h]
\caption{Genetic Algorithm for Optimal Selection of EEP or EEP Combinations}\label{alg:genetic_algorithm}
\begin{algorithmic}[1]
    \REQUIRE Population size $N$, Number of generations $G$, Crossover probability $P_c$, Mutation probability $P_m$
    \STATE Initialize a random population of size $N$
    \FOR{$g = 1$ to $G$}
        \STATE Evaluate the fitness of each individual in the population
        \STATE Sort the population based on fitness in descending order
        \STATE Create an empty next-generation population
        \WHILE{next-generation population is not full}
            \STATE Select two parents based on fitness
            \IF{random number $< P_c$}
                \STATE Perform crossover to create two offspring
            \ELSE
                \STATE Copy the parents to the next-generation population
            \ENDIF
            \FOR{each offspring}
                \IF{random number $< P_m$}
                    \STATE Perform mutation on the offspring
                \ENDIF
                \STATE Add the offspring to the next-generation population
            \ENDFOR
        \ENDWHILE
        \STATE Replace the current population with the next-generation population
    \ENDFOR
    \RETURN The best EEP or EEP combination in the final population is the solution
\end{algorithmic}
\end{algorithm}

\begin{table*}[ht]
\caption{Performance comparison of the proposed multistage classifier and other approaches.}
\centering
\setlength{\tabcolsep}{3pt}
\resizebox{\linewidth}{!}{
\begin{tabular}{@{}lcclcccc@{}}
\toprule
\textbf{Authors} & \textbf{Network} &\textbf{Dataset}&\textbf{Methods} & \textbf{Performance \& Complexity}\\ \midrule
Murugesan~\emph{et~al.}~\cite{murugesan2018ecgnet}&CNN&MIT-BIH&End-to-end, 720 input samples &Accuracy 97.6\%, FLOPs 12718080$^*$\\
Hannun~\emph{et~al.}~\cite{hannun2019cardiologist}  &DNN & Proprietary &End-to-end, 256 input samples, 12 classes  & Sensitivity 83.7\%, FLOPs 20971520$^*$\\
Xia~\emph{et~al.}~\cite{xia2019novel} &CNN&MIT-BIH&End-to-end, 200 input samples&Accuracy 99.24\%, Sensitivity 98.78\%, FLOPs 2578640 \\
Huang~\emph{et~al.}~\cite{huang2023novel}&CNN-LSTM &MIT-BIH& Time representation input &Accuracy 98.95\%, Sensitivity 96.54\%, FLOPs 26417728\\
Plawiak~\cite{plawiak2018novel}  &SVM & MIT-BIH & 3600 samples (10 s)  & Accuracy 98.85\%, Sensitivity 90.2\%, FLOPs 249000\\
Ozal~\emph{et~al.}~\cite{17classes} & CNN  &MIT-BIH &ECG fragment (10 s)    & Accuracy 91.33\%, Sensitivity 83.91\%, FLOPs 1665904\\
Kiranyaz~\emph{et~al.}~\cite{7202837}&CNN&MIT-BIH&End-to-end, different beat representations &Accuracy 97.5\%, Sensitivity 79.8\%, FLOPs 1859320$^*$\\
Muhammad~\emph{et~al.}~\cite{safdar2023novel} &CNN &PTB-XL~\cite{wagner2020ptb} &Image Input, 12-ECG leads &Accuracy 89.87\%, Sensitivity 88.99\%, FLOPs 48143841$^*$\\
Oh~\emph{et~al.}~\cite{oh2018automated}  & CNN-LSTM &MIT-BIH &Variable lengths   &Accuracy 98.10\%, Sensitivity 97.50\%, FLOPs 79460338\\
\midrule
\textbf{Original work}&CNN & MIT-BIH &End-to-end &Accuracy 98.41\%, Sensitivity 99.14\%, FLOPs 303104\\ 
\multirow{6}{*}{\textbf{This work (Ours)}} & \multirow{6}{*}{CNN} & \multirow{6}{*}{MIT-BIH}   & DNN Partitioning, EEP 2 (thre=0.8) &Accuracy 97.68\%, Sensitivity 98.16\%, FLOPs 232780\\  
       & & & DNN Partitioning, EEP 3 (thre=0.8) &Accuracy 97.56\%, Sensitivity 98.55\%, FLOPs 239093\\
       & & & DNN Partitioning, EEP 4 (thre=0.8) &Accuracy 97.56\%, Sensitivity 98.40\%, FLOPs 262214\\ 
       & & & DNN Partitioning, EEP 1, EEP 3 (thre=0.9) &Accuracy 96.73\%, Sensitivity 98.58\%, FLOPs 232885\\  
       & & & DNN Partitioning, EEP 2, EEP 4 (thre=0.9) &Accuracy 97.74\%, Sensitivity 98.36\%, FLOPs 236629\\
       & & & DNN Partitioning, EEP 3, EEP 5 (thre=0.9) &Accuracy 97.85\%, Sensitivity 98.61\%, FLOPs 243928\\ 
       \bottomrule
\end{tabular}
}
\label{tbl:literature_comparison}
\vspace{0.5\baselineskip} 
\footnotesize 
\begin{flushleft} 
    *This value is an estimate based on assumptions over the operations involved and architectures unveiled.
\end{flushleft}
\end{table*}

Table~\ref{tbl:literature_comparison} lists several approaches for ECG classification and compares those with the method we proposed. We made some assumptions when calculating the FLOPs of algorithms that lack detailed architectural information. For algorithms with detailed architectures, we calculate based on the architecture. Notably, the complexity of traditional ML models is relatively low compared to DL models. While SVM models are among the more complex within ML, their sensitivity remains lower, falling short of the performance benchmarks set by more recent models~\cite{plawiak2018novel}. Many existing studies utilize large-scale DNNs, requiring significant computations (FLOPs) to achieve relatively high performance. However, these models lack the capability to be distributed on different nodes.
\cite{17classes} employed 16-layer CNN to classify 10-second non-overlapping samples into 17 classes, achieving an accuracy of 91.33\%. However, the server-deployed approach to achieve high performance imposes challenges on network bandwidth, quality and reliability.
\cite{oh2018automated} first utilized variable-length signals for ECG classification, achieving 98.10\% accuracy. however, its computational intensity necessitates reliance on cloud servers, which can hinder performance during network interruptions.
In contrast, DCenNet features a decentralized multistage inferencing system that distributes computations across nodes, effectively addressing these limitations. This capability not only enhances performance and reliability but also ensures timely responses to abnormal signals, making it a more robust choice for ECG monitoring.

\subsection{Single EEP}

Adding EEPs into a large CNN model enables a distributed inferencing system that facilitates the integration of the model into various decentralised practical systems.
Through empirical analysis, we achieved a well-balanced solution that ensures optimal performance and resource efficiency, showcasing the adaptability of DCenNet for AI inferencing in the Edge-Cloud continuum. 

Fig.~\ref{fig:exitPerformance_TwoParts} illustrates the variations in performance and complexity as the confidence threshold changes for different locations of the single EEP. The performance superiority of the final exit point over the initial exit point is a common observation in systems employing a single EEP for inference. However, the correlation between the placement of the EEP and the resultant performance is not always straightforward. While intuitively, one might expect a later EEP to yield better outcomes, empirical evidence suggests that this isn't always the case. Fig.~\ref{fig:exitPerformance_TwoParts} shows that the system accuracy is higher with EEP 2 compared to either EEP 3 or EEP 4, and the system sensitivity is consistently higher with EEP 4 than with EEP 5. 
In systems featuring only one EEP, opting for an earlier exit point undoubtedly reduces the FLOPs required for computation. Conversely, selecting a later exit point does not always lead to enhanced performance metrics. It would appear that delaying the early exit point would result in superior system performance despite the potential increase in FLOPs. However, empirical evidence suggests that this isn't consistently valid. This suggests that factors beyond FLOPs, such as the architecture of the model, the nature of the dataset, and the specific task being performed, play crucial roles in determining the optimal placement of exit points. It can be observed from Fig.~\ref{fig:exitPerformance_TwoParts} that if exiting at EEP 2, while ensuring a relatively high level of performance, selecting a threshold of 0.8 could result in approximately $\sim \!\! 24\%$ reduction in FLOPs consumption. Similarly, with an early exit at EEP 3, choosing a threshold of 0.8 would ensure both high accuracy and relatively high sensitivity while achieving a $\sim \!\! 21\%$ reduction in FLOPs usage. Because the relationship between the position of EEP and system performance is complex, it underscores the importance of the system design we proposed.

\subsection{Two EEPs}
We expanded our ECG classification model by incorporating two additional EEPs, thus dividing the original model into three subnetworks. This modification allows for the deployment of the model across three decentralized nodes within the Edge-Fog-Cloud continuum. We conducted experiments on the model with two EEPs, exploring various combinations, and the performance of the proposed system is shown in Fig.~\ref{fig:exitPerformance_ThreeParts} at varying confidence thresholds. Exit rate 1 represents the proportion of outcomes exceeding the confidence threshold at the first EEP, while exit rate 2 represents the proportion of outcomes exceeding the confidence threshold at the other EEP, both considered reliable relative to the overall count.
Introducing two EEPs into the model, especially when the first EEP is positioned after layer 1, presents an enhanced capacity to reduce overall FLOPs. However, this advantage comes at the cost of a noticeable decrease in system accuracy. Placing the first EEP after the initial convolutional layer necessitates maintaining an exceptionally high confidence threshold to attain a satisfactory level of accuracy. Despite the evident advantage of FLOPs conservation through EEP utilization, the system's accuracy experiences a discernible 5\% decline compared to alternative strategies employing a threshold of 0.5.
When opting for the combination of EEPs 3 and 4, with a confidence threshold set at 0.8, the system achieves an impressive sensitivity of 98.4\% and an accuracy of 97\%. This configuration also leads to a substantial reduction of $\sim \!\! 20\%$ in FLOPs consumption. 
Based on the result in Fig.~\ref{fig:exitPerformance_ThreeParts}, a lower threshold makes it easier to exit at the first EEP, leading to a higher Exit Rate 1 when the threshold is low; conversely, a higher threshold makes it more difficult to exit at the first EEP, making it more likely to exit at the other EEP or the final EP, hence Exit Rate 2 increases when the threshold goes high. As the threshold approaches 1, signals that are difficult to classify will exit at the final EP, causing Exit Rate 2 to decrease slightly when the threshold approaches 1.
These findings underscore the complex trade-offs involved in selecting the positioning and threshold values for multiple EEPs, highlighting the importance of carefully designing such systems to balance computational efficiency and performance metrics.

\subsection{Optimization Method: Genetic Algorithm (GA)}
Achieving high accuracy and sensitivity alongside low complexity simultaneously is not feasible, hence it is necessary to identify a trade-off based on the specific requirements at hand. GA is an evolutionary optimization algorithm inspired by the mechanisms of natural selection and genetics, which has gained significant popularity for solving complex optimization problems across various domains~\cite{geneticALG1,geneticALG2}. GA is particularly suited for optimizing the placement of EEPs in decentralized systems, as it balances competing goals such as accuracy, sensitivity, and computational complexity. GA starts with a population of potential solutions represented as strings of information, shown in Algorithm~\ref{alg:genetic_algorithm}. Through selection, crossover (mixing solutions), and mutation (changing them slightly), new generations of solutions are produced. This allows GA to explore and improve solutions over time. The fitness of each candidate solution is evaluated based on an objective function (OF) shown in Eq.~\eqref{eq:fitness}, guiding the algorithm towards finding optimal or near-optimal solutions. Normalization of each parameter (accuracy, sensitivity, and FLOPs) is essential in this process. Once the weight for each parameter is assigned based on the actual requirements, the GA will return the corresponding exit points and related performance.

\begin{alignat}{2}
    \textit{OF} &=&~& w_{acc} \times \textit{accuracy} \label{eq:fitness} \\
                     & &+& w_{sen} \times \textit{sensitivity} \nonumber \\
                     & &-& w_{com} \times \textit{FLOPs} \nonumber
\end{alignat}
where, $w_{acc}$, $w_{sen}$, and $w_{com}$ represent the weight of accuracy, sensitivity, and the number of FLOPs for the OF.

By assigning varied values to these weights, taking into account specific practical considerations and performance requirements, we can effectively determine the optimal EEP or combinations of EEPs. This method provides a flexible approach for balancing competing objectives, and its adaptability is critical for the system, where wearable devices must meet performance constraints while maintaining energy efficiency. When all weights are set to 1, according to the results provided by GA, the optimal EEP is EEP 2 when only one EEP is introduced in the system. With this point, the system's accuracy is 96.68\%, sensitivity is 97.73\%, and the total FLOPs is 230.33k. 
For two EEPs applied Edge-fog-cloud system, the optimal combination of EEPs is EEP 2 and EEP 5 with equal weights of $w_{acc}$, $w_{sen}$, and $w_{com}$, achieving an accuracy of 96.21\%, sensitivity of 97.62\%, and a total FLOPs of 229.34k. When $w_{acc}$ and $w_{sen}$ are set to 10, and $w_{com}$ is 1, indicating a probability to increase FLOPs for better model performance, the optimal EEPs combination becomes EEP 3 and EEP 5, resulting in an accuracy of 95.81\%, sensitivity of 97.84\%, and total FLOPs of 235.4k. Lastly, when $w_{acc}$ and $w_{sen}$ are set to 1 each, and $w_{com}$ is 100, showing a probability to sacrifice metrics for complexity reduction, the optimal EEPs combination is to EEP 1 and EEP 5, resulting in an accuracy of 89.89\%, sensitivity of 97.11\%, and total FLOPs of 220.42k.

This methodology allows us to find a trade-off between essential elements such as accuracy, sensitivity, and computational efficiency, thus enabling us to select the optimal trade-off. Ensuring that the weights are normalized is vital to ensure fair and precise comparisons across various parameter configurations. This method helps us figure out the best way to make models better so they fit what we need them to do in the real world.

\subsection{Power Consumption Analysis}

We tested 500 ECG heart beats for each threshold level, with the heartbeats evenly distributed across the 5 classes. The threshold was set between 0.5 and 0.9 because thresholds below 0.5 yielded classifications with low confidence. To minimize interference and ensure consistent power consumption data, all non-essential onboard devices, including sensors, LEDs, and unused peripherals, were deactivated during the experiments.



\begin{figure}[!htb]
    \centering
    \includegraphics[width=\linewidth]{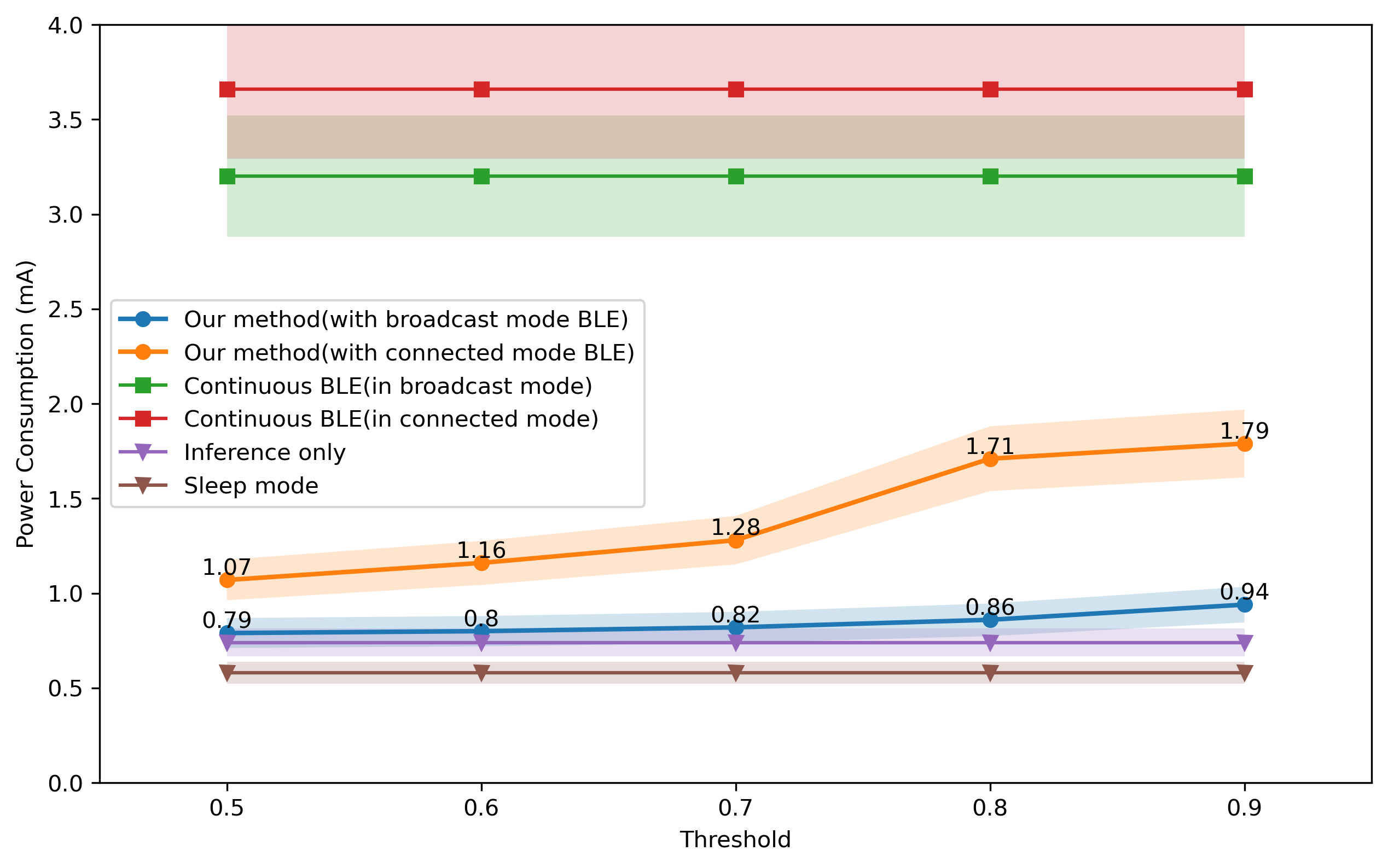}
    \caption{The power consumption comparison of continuous BLE (two modes), our method (two modes), inference-only mode, and sleep mode.}
    \label{fig:Power}
\end{figure}

We tested 6 different configuration modes of power consumption for better comparison. Fig.~\ref{fig:Power} illustrates the upper bound (i.e., the power consumption during continuous BLE transmission) and the lower bound (i.e., the power consumption in sleep mode). Our best model, which utilizes a lightweight CNN classifier and applies BLE in broadcast mode, can on average save more than 73.6\% of power compared to the best method of continuous BLE ECG signal transmission, and on average, it consumes only 13.8\% more power than the normal calculation mode (i.e. inference-only mode).

%% file: Sections/6_conclusion.tex
\section{Conclusion}
\label{conclusion}
In this paper, we proposed DCenNet, a decentralization approach for biomedical signals classification using large cloud-centralized networks by introducing one or two EEPs. We demonstrated that adding a single EEP allows the partitioning of the network into an Edge-Cloud continuum while deploying two EEPs creates an Edge-Fog-Cloud continuum, effectively dividing the large network into three parts. This design not only enhances the responsiveness to abnormal signals but also reduces power consumption at the edge, making it an economically efficient solution. Also, by moving the EEPs to different locations in the model, trade-offs among model performance, resource utilization, and complexity must be achieved. The Edge-Cloud continuum exhibited an accuracy of 97.56\% and a sensitivity of 98.55\%, accompanied by a notable $\sim \!\! 21\%$ reduction in complexity. Similarly, the Edge-Fog-Cloud continuum demonstrated an accuracy of 97.74\% and a sensitivity of 98.36\%, achieving a $\sim \!\! 22\%$ reduction in FLOPs. These outcomes serve as strong validation for the efficacy of our proposed decentralized multistage system design. The earlier the EEPs are deployed, the greater the reduction in the number of FLOPs, but deploying them later does not necessarily correspond to an increase in system performance. By increasing the confidence threshold to a certain extent, both continuums can achieve the performance of the original model, and even surpass its capabilities. In the future, we can incorporate varying numbers of EEP for different system requirements to enhance the adaptability of the model.